\documentclass[aps,print,prb,twocolumn,superscriptaddress]{revtex4-2}

\usepackage[hidelinks]{hyperref}
\hypersetup{
	colorlinks,
	linkcolor={red},
	citecolor={blue},
	urlcolor={blue}
}

\usepackage{balance}
\usepackage{amssymb}
\usepackage{suffix}
\usepackage{mathtools}
\usepackage[utf8]{inputenc}
\usepackage{booktabs}
\usepackage{cases}
\usepackage[multiple]{footmisc}
\usepackage{dcolumn}
\usepackage{color,soul}
\usepackage{rotating}
\usepackage{perpage}
\usepackage{siunitx}
\usepackage{xcolor}
\usepackage{soul}
\usepackage{amsmath}
\usepackage{tikz}
\usepackage[T1]{fontenc}
\usepackage{etoolbox}
\usepackage{graphics}
\usepackage{siunitx}
\usepackage{float}	
\usepackage{collref}
\usepackage{multirow}
\usepackage{mathtools}
\usepackage{bm}
\usepackage{url}

\usepackage{tikz}
\usepackage{tikz-3dplot}
\usepackage{accents}

\makeatletter
\newcommand{\doublewidetilde}[1]{{%
		\mathpalette\double@widetilde{#1}%
}}
\newcommand{\double@widetilde}[2]{%
	\sbox\z@{$\m@th#1\widetilde{#2}$}%
	\ht\z@=.9\ht\z@
	\widetilde{\box\z@}%
}
\makeatother

\usepackage{etoolbox,lipsum}

\begin{document} 

	\title{RKKY interaction in altermagnets with adiabatic electron-phonon coupling}
	\author{Bui D. Hoi}
	\email{buidinhhoi@hueuni.edu.vn}
	\address{Faculty of Physics, University of Education, Hue University, Hue 530000, Vietnam} 
	
	\date{\today}
	
	\begin{abstract}
		Altermagnets, characterized by time-reversal symmetry breaking without net magnetization and momentum-dependent spin-split bands, offer a promising platform for spintronics due to their anisotropic spin textures and potential for tunable magnetic interactions. Here, we theoretically investigate the slow phonon-renormalized Ruderman-Kittel-Kasuya-Yosida (RKKY) interaction in two-dimensional Rashba $d$-wave altermagnets, incorporating arbitrary Rashba spin-orbit coupling (RSOC) strength and spin-dependent static-Holstein electron-phonon coupling (EPC). Using a continuum model Hamiltonian that captures altermagnetic anisotropy, RSOC, and adiabatic lattice distortions, we compute the noncollinear RKKY exchange tensor via second-order perturbation theory and numerical momentum-space integration. Our results reveal that static lattice distortions provide a versatile tuning knob for engineering the magnitude, anisotropy, and chirality of RKKY couplings: moderate EPC suppresses long-range coherence while inducing component-specific phase shifts and sign reversals, particularly in Dzyaloshinskii-Moriya and compass-like terms, enabling control over ferromagnetic/antiferromagnetic alignments and clockwise/counterclockwise chiralities. Systematic parameter scans demonstrate enhanced oscillatory complexity with altermagnetic order, RSOC, and doping. These findings establish slow phonons as a low-energy tuning knob that systematically controls the magnitude, anisotropy, phase, and chirality of noncollinear RKKY couplings—even at arbitrary Rashba strength—thereby providing a practical route to engineer ferromagnetic/antiferromagnetic alignments and clockwise/counterclockwise DM chiralities in altermagnet-based heterostructures.
	\end{abstract}
	
	\maketitle
	{\allowdisplaybreaks
		
		
		\section{Introduction}
		Magnetic materials hosting unconventional spin textures have long been at the center of research in condensed matter physics due to their potential to advance spin-based electronics~\cite{Fert2017,Jungwirth2016,RevModPhys.76.323,RevModPhys.90.015005}. Among these, altermagnets have recently emerged as a unique platform where time-reversal symmetry is broken while preserving spatial inversion, leading to compensated magnetic ordering and momentum-dependent spin-split electronic bands~\cite{PhysRevX.12.040501,PhysRevX.12.040002,PhysRevX.12.031042,doi:10.7566/JPSJ.88.123702,PhysRevX.12.011028,PhysRevB.102.014422,PhysRevB.99.184432,PhysRevMaterials.5.014409,he2025altermagnetismttprimedeltafermihubbardmodel,7bss-9yxb}. These systems host vanishing macroscopic magnetization yet exhibit spin textures with pronounced anisotropy, opening opportunities for novel spintronic functionalities~\cite{PhysRevLett.130.216701,PhysRevLett.128.197202,PhysRevLett.129.137201,PhysRevLett.134.106801,PhysRevLett.134.106802,PhysRevB.110.235101,PhysRevLett.131.076003,PhysRevB.108.054511,PhysRevB.108.L060508,PhysRevB.108.184505,PhysRevB.108.224421,Zhang2024,PhysRevB.109.224502,PhysRevB.108.205410}. The interplay of symmetry and magnetism in altermagnets produces highly anisotropic spin structures, often described by \( d \)-, \( g \)-, or \( f \)-wave form factors~\cite{doi:10.1073/pnas.2108924118,PhysRevLett.130.216701,PhysRevLett.132.086701,PhysRevLett.132.176702}, enabling strongly spin-polarized states in the absence of net magnetization. Experimental signatures of altermagnetic phases have been reported in materials such as RuO\(_2\)~\cite{Feng2022,doi:10.1126/sciadv.aaz8809,PhysRevLett.128.197202,weber2024opticalexcitationspinpolarization} and MnTe~\cite{PhysRevLett.132.036702,PhysRevB.107.L100418,PhysRevB.107.L100417}, while theoretical studies have proposed candidate compounds including La\(_2\)CuO\(_4\)~\cite{PhysRevX.12.031042}, MnF\(_2\)~\cite{PhysRevB.102.014422}, and Mn\(_5\)Si\(_3\)~\cite{PhysRevX.12.040501,reichlova2021macroscopictimereversalsymmetry}. 
		
		A central avenue for exploiting the spin-dependent electronic structure of altermagnets is through the Ruderman--Kittel--Kasuya--Yosida (RKKY) interaction, where magnetic impurities communicate via itinerant electrons~\cite{10.1143/PTP.16.45,*PhysRev.106.893,*PhysRev.96.99}. In altermagnets, the oscillatory and anisotropic character of this interaction acquires additional complexity due to the momentum-dependent spin splitting, as demonstrated in recent studies~\cite{lee2024magneticimpuritiesaltermagneticmetal,PhysRevB.110.054427,k3xb-8pts,kundu2025rkkyinteractionmediatedspinpolarized,yarmohammadi2026efficienttwocolorfloquetcontrol,gr5n-174c}. Moreover, external tuning mechanisms such as magnetic fields, gate-controlled Rashba spin-orbit coupling~(RSOC), or optical driving have been proposed to manipulate the RKKY coupling strength and symmetry properties, offering promising routes toward dynamical control of spin interactions for spintronic and quantum-information applications.  
		
		Recent studies have extensively explored the RKKY interaction in altermagnets. In the weak RSOC regime, where analytical progress is feasible, Amundsen~\textit{et al.}~\cite{PhysRevB.110.054427} and Yarmohammadi~\textit{et al.}~\cite{k3xb-8pts} investigated the RKKY interaction. Amundsen~\textit{et al.}~\cite{PhysRevB.110.054427} demonstrated that the momentum-resolved spin polarization of itinerant carriers in altermagnets qualitatively modifies the RKKY interaction compared to the isotropic spin splitting in ferromagnets. Yarmohammadi~\textit{et al.}~\cite{k3xb-8pts} further showed that the intrinsic strength of altermagnetism imprints chirp-like signatures in the magnetic response, which can be dynamically tuned via optical driving.
		In the regime of stronger RSOC, where numerical approaches are required, Kundu~\cite{kundu2025rkkyinteractionmediatedspinpolarized}, Yarmohammadi~\textit{et al.}~\cite{yarmohammadi2026efficienttwocolorfloquetcontrol}, and Duan~\textit{et al.}~\cite{gr5n-174c} have examined the interaction. Kundu~\cite{kundu2025rkkyinteractionmediatedspinpolarized} reported that the Dzyaloshinskii-Moriya~(DM) vector lies in the two-dimensional~(2D) plane, with the DM interaction being odd in RSOC strength, oscillatory, and increasing in magnitude with RSOC.  Very recently, Yarmohammadi~\textit{et al.}~\cite{yarmohammadi2026efficienttwocolorfloquetcontrol} demonstrated that two-color laser driving provides efficient and tunable control over all components of the RKKY interaction using two weak laser fields. Duan~\textit{et al.}~\cite{gr5n-174c} found that the Néel vector orientation can be accurately determined from the Ising term in the absence of SOC, or qualitatively inferred via DM terms when SOC is present. 
		
		Electron-phonon coupling (EPC) is known to renormalize quasiparticle dispersions, influence spin transport, and mediate superconductivity in correlated systems~\cite{giustino2016electron,6wxh-p4mc,leraand2025phononmediatedspinpolarizedsuperconductivityaltermagnets,nlpj-1dt5,hodt2025phononenhancedopticalspinconductivityspinsplitter}. In altermagnets, EPC has been shown to hybridize collective modes~\cite{nlpj-1dt5}, enhance spin conductivity~\cite{hodt2025phononenhancedopticalspinconductivityspinsplitter}, stabilize unconventional superconducting states~\cite{6wxh-p4mc,leraand2025phononmediatedspinpolarizedsuperconductivityaltermagnets}, and control spin Edelstein (de)polarization~\cite{yarmohammadi2025slowphononcontrolspinedelstein}. However, its role in shaping magnetic impurity interactions—particularly how slow phonons affect the oscillatory and noncollinear RKKY exchange in Rashba 
		$d$-wave altermagnets with arbitrary RSOC strength—remains largely unexplored. Addressing arbitrary RSOC strength is crucial to capturing anisotropic exchange contributions that may be missed in the weak-RSOC limit~\cite{kundu2025rkkyinteractionmediatedspinpolarized,gr5n-174c,yarmohammadi2026efficienttwocolorfloquetcontrol}.
		
		In this work, we address this gap by systematically analyzing the phonon-driven noncollinear RKKY interaction in 2D Rashba $d$-wave altermagnets. Our study is motivated by recent work~\cite{yarmohammadi2025slowphononcontrolspinedelstein}, which employed a continuum model incorporating RSOC and spinful static-Holstein-type EPC to calculate the spin Edelstein polarization and clarify the interplay between altermagnetic order and EPC. Here, we extend this framework to compute the impurity-exchange interaction in the presence of the same slow lattice vibrations. Throughout this work we treat the electron-phonon interaction in the adiabatic (static-Holstein) limit appropriate for slow optical phonons. All results are obtained at zero temperature. Treating EPC at the static-Holstein level while retaining the full angular dependence of the RKKY kernel allows us to identify regimes in which phonon renormalization qualitatively modifies spin-mediated impurity couplings. By systematically mapping the dependence on phonon coupling strength, Rashba interaction, and doping, we demonstrate that slow phonons provide a tunable handle to control both the magnitude and anisotropy of RKKY interactions in altermagnets.
		
		The remainder of this paper is organized as follows. In Sec.~\ref{s2}, we introduce the model Hamiltonian including slow-phonon coupling. We also detail the methodology for evaluating the phonon-renormalized RKKY interaction. Results and discussions are presented in Sec.~\ref{s3}, followed by concluding remarks in Sec.~\ref{s4}.  

		
\section{Theoretical framework}\label{s2}
To investigate the indirect exchange interaction between localized magnetic moments in a 2D altermagnet with momentum-dependent spin splitting, we employ a continuum model formulated near the band minimum. The electronic Hamiltonian incorporates a parabolic dispersion combined with an angularly dependent spin-splitting term, as well as an additional contribution originating from structural inversion asymmetry that couples the orbital motion to the spin degrees of freedom.

\subsection{Effective Hamiltonian}
The low-energy quasiparticles of a Rashba altermagnet can be described by the unperturbed Hamiltonian~\cite{winkler2003spin}
\begin{equation}\label{eq:H_rashba_altermagnet}
	H_{\mathbf{q}} = \epsilon_{\mathbf{q}} \, I + \delta_{\mathbf{q}} \, \tau_z + \alpha \left(q_y \, \tau_x - q_x \, \tau_y\right),
\end{equation}
where \(I\) is the identity operator in spin space, and \(\tau_i\) (\(i=x,y,z\)) are the standard Pauli matrices representing spin degrees of freedom. The first term, \(\epsilon_{\mathbf{q}} =  \frac{\hbar^2 q^2}{2 m_e}\), describes the isotropic kinetic energy of carriers, which preserves fourfold rotational symmetry in the plane. The second term, $\delta_{\mathbf{q}} = \frac{\hbar^2 \delta}{2 m_e} (q_x^2 - q_y^2)$ introduces a momentum-dependent spin splitting along \(\tau_z\), parametrized by \(0<\delta<1\), capturing the characteristic anisotropy of the altermagnetic exchange. Finally, the third term represents the RSOC
\(\alpha (q_y \tau_x - q_x \tau_y)\), where \(\alpha\) quantifies the strength of spin-momentum entanglement induced by inversion asymmetry. 

The combination of these terms results in a Hamiltonian in which the spin eigenstates are intertwined with the crystal momentum, while the kinetic part preserves the underlying lattice symmetry. The directional variation introduced by \(\delta_{\mathbf{q}}\) breaks spin degeneracy along orthogonal momentum directions, providing a natural mechanism for anisotropic spin splitting characteristic of altermagnetic Rashba systems~\cite{k3xb-8pts,yarmohammadi2025slowphononcontrolspinedelstein,yarmohammadi2025spinpolarizationengineeringdwave}.

It should be noted that here we focus on a $d_{x^2-y^2}$-wave altermagnet; however, the formalism can be straightforwardly extended to $d_{xy}$-wave or other symmetry classes. While the qualitative and quantitative behavior of the system may vary depending on the specific wave pattern, the formulation of static lattice distortion effects remains general and can be applied to other altermagnetic configurations as well.\begin{figure}[t]
	\centering
	\includegraphics[width=1\linewidth]{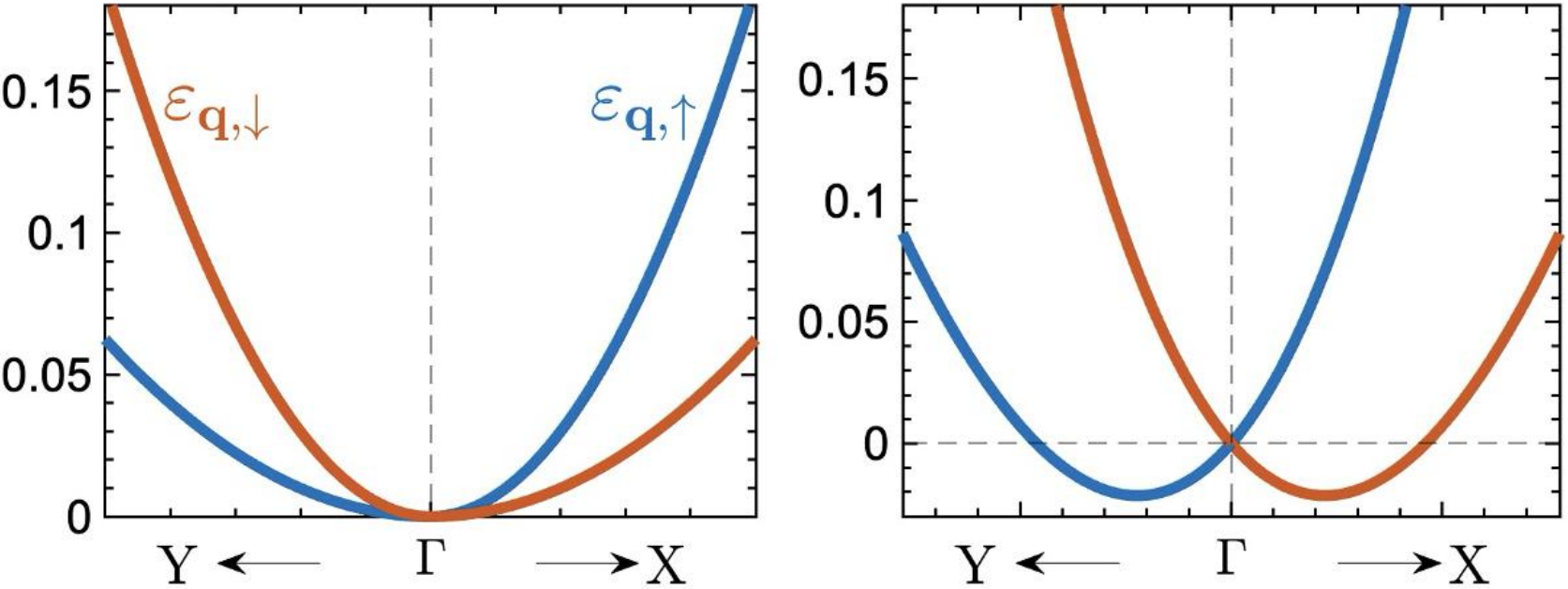}
	\caption{Low-energy electronic dispersion, i.e., the band structure, along the $Y \leftarrow \Gamma \rightarrow X$ high-symmetry path of the first Brillouin zone for a 2D Rashba $d_{x^2-y^2}$-wave altermagnet described by Eq.~\eqref{eq_2}, shown in the absence (left panel) and in the presence (right panel) of RSOC $\alpha = 0.5$ eV$\cdot$\AA.}
	\label{f1}
\end{figure}  

The spectral properties of the Rashba altermagnet are obtained by diagonalizing the Hamiltonian in Eq.~\eqref{eq:H_rashba_altermagnet}, yielding the energy eigenvalues
\begin{equation}\label{eq_2}
	\varepsilon_{\mathbf{q},s} = \epsilon_{\mathbf{q}} + s \sqrt{\delta_{\mathbf{q}}^2 + \alpha^2 q^2},
\end{equation}
where \(s = \pm 1\) labels the two spin-split branches, as shown along the $Y \leftarrow \Gamma \rightarrow X$ path of the first Brillouin zone in Fig.~\ref{f1} without (left) and with (right) RSOC, and \(q = |\mathbf{q}|\) is the magnitude of the in-plane momentum. The first term, \(\epsilon_{\mathbf{q}}\), provides the isotropic kinetic contribution, while the square-root term encodes the combined effect of the altermagnetic anisotropy, \(\delta_{\mathbf{q}}\), and the RSOC, \(\alpha\).  The corresponding eigenstates are two-component spinors oriented either parallel or antiparallel to the effective momentum-dependent field
\(\mathbf{h}_{\mathbf{q}}^{\rm eff} = (\alpha q_y, -\alpha q_x, \delta_{\mathbf{q}})\). Consequently, the spin expectation value of each eigenstate is not aligned along a fixed axis but rather follows the local direction of \(\mathbf{h}_{\mathbf{q}}^{\rm eff}\), reflecting the entanglement of spin and momentum in the system. This structure is responsible for the anisotropic spin splitting and underlies the peculiar spin-dependent transport.

Next, we include lattice vibrational degrees of freedom, modeled as site-local harmonic oscillators that modulate the local electronic environment and can couple differently to opposite spin projections~\cite{yarmohammadi2025slowphononcontrolspinedelstein}. The bosonic Hamiltonian is given by
\begin{equation}
	H_{\rm vib} = \sum_j \left( \frac{P_j^2}{2 M_j} + \frac{1}{2} M_j \omega_v^2 X_j^2 \right),
\end{equation}
where \(X_j\) represents the displacement of the vibrational mode at site \(j\), \(P_j\) is the corresponding canonical momentum, \(M_j\) is the ionic mass, and \(\omega_v\) denotes the vibrational mode frequency. This term describes a set of independent harmonic oscillators at each lattice site.  

The coupling between electrons and vibrations is taken as
\begin{equation}
	H_{\rm int} = \sum_j \left[ \lambda_+ \left( d_{j,\uparrow} - \frac{1}{2} \right) + \lambda_- \left( d_{j,\downarrow} - \frac{1}{2} \right) \right] X_j,
\end{equation}
where \(d_{j,\sigma}\) counts the number of electrons with spin projection \(\sigma\) at site \(j\), and \(\lambda_\pm\) denote the spin-dependent coupling strengths. The subtraction of \(1/2\) ensures that the coupling is referenced relative to half-filling, such that the vibrational displacement couples to spin polarization rather than to the total charge.  

Introducing the average coupling, \(\lambda = (\lambda_+ + \lambda_-)/2\), and the difference, \(\tilde{\lambda} = (\lambda_+ - \lambda_-)/2\), naturally separates the electron-phonon interaction into charge and spin channels:
\begin{equation}
	H_{\rm int} = \sum_j \lambda \left( n_j - 1 \right) X_j + \sum_j \tilde{\lambda} \, S_j^z X_j,
\end{equation}
where \(n_j = d_{j,\uparrow} + d_{j,\downarrow}\) is the local electron density and \(S_j^z = d_{j,\uparrow} - d_{j,\downarrow}\) is the spin polarization. This decomposition explicitly shows that the vibrational modes can mediate both conventional charge coupling and spin-selective interactions, which is particularly relevant for generating spin-dependent nonlinearities in the altermagnetic lattice.

For vibrational modes that are slow compared to the electronic dynamics, the ionic displacement can be treated quasi-statically. In this adiabatic approximation, we replace the local deviation by its average value, \(\langle X_j \rangle = X\), leading to an effective electronic Hamiltonian
\begin{equation}
	H_{\mathbf{q}}^{\rm eff} = (\epsilon_{\mathbf{q}} + \lambda X) \, I + (\delta_{\mathbf{q}} + \tilde{\lambda} X) \, \tau_z + \alpha \left( q_y \, \tau_x - q_x \, \tau_y \right),
\end{equation}
where the EPC terms \(\lambda X\) and \(\tilde{\lambda} X\) shift the charge and spin channels, respectively. This Hamiltonian effectively captures the feedback of the slow lattice on the electronic quasiparticles. The steady-state displacement \(X\) is determined by minimizing the total energy, balancing the lattice restoring force against the electronic response $M_j \omega_v^2 X = - \lambda (n_t - 1) - \tilde{\lambda} \, \mathcal{S}$, where \(n_t = d_\uparrow + d_\downarrow\) is the total electron occupancy per site, and \(\mathcal{S} = d_\uparrow - d_\downarrow\) is the spin polarization. These quantities are obtained from the electronic states of the effective Hamiltonian as
\begin{align}
	n_t &= \frac{1}{N_q} \sum_{\mathbf{q},s} f(\varepsilon_{\mathbf{q},s}^{\rm eff} - \varepsilon_{\rm F}), \\
	p &= \frac{1}{N_q} \sum_{\mathbf{q},s} s \, \langle \phi_{\mathbf{q},s} | \tau_z | \phi_{\mathbf{q},s} \rangle \, f(\varepsilon_{\mathbf{q},s}^{\rm eff} - \varepsilon_{\rm F}),
\end{align}
where \(f(x) = \left[ \exp(x/k_B T) + 1 \right]^{-1}\) is the Fermi--Dirac distribution at temperature \(T\), \(\varepsilon_{\rm F}\) is the chemical potential, and \(N_q\) is the number of momentum points in the Brillouin zone. The energies entering these expressions are the adiabatically adjusted eigenvalues $\varepsilon_{\mathbf{q},s}^{\rm eff} = \epsilon_{\mathbf{q}} + \lambda X + s \sqrt{ (\gamma_{\mathbf{q}} + \tilde{\lambda} X)^2 + \alpha^2 q^2 }$, with \(|\phi_{\mathbf{q},s}\rangle\) the corresponding eigenstates of \(H_{\mathbf{q}}^{\rm eff}\).  

These relations must be solved self-consistently: starting from an initial guess for \(X\), one computes \(n_t\) and \(\mathcal{S}\), updates \(X\) via the lattice balance equation, and iterates until convergence. This procedure yields the equilibrium lattice displacement and the renormalized electronic spectrum for a given set of couplings \((\lambda, \tilde{\lambda})\), temperature \(T\), and chemical potential \(\varepsilon_{\rm F}\). In the present study, we focus on the zero-temperature limit, where the electronic states are occupied up to the Fermi energy and thermal fluctuations are absent. The self-consistent feedback encodes the interplay between slow ionic motion and spin-dependent electronic structure, a key ingredient for understanding spin-selective responses in altermagnetic materials.

\subsection{RKKY interaction}
To quantify the indirect exchange interaction between two localized magnetic moments, \(\mathbf{M}_1\) at position \(\mathbf{R}_1\) and \(\mathbf{M}_2\) at \(\mathbf{R}_2\), separated by \(\mathbf{R} = \mathbf{R}_2 - \mathbf{R}_1\), we employ second-order perturbation theory in the local contact exchange interaction,  
\begin{equation}
	H_{\rm int}^{(l)} = \mathcal{J}\, \mathbf{M}_l \cdot \boldsymbol{\tau}, \quad l = 1,2,
\end{equation}  
where $\mathcal{J}$ denotes the local exchange coupling constant and \(\boldsymbol{\tau} = (\tau_x, \tau_y, \tau_z)\) shows the Pauli matrices acting on the electronic spin. At second order, the effective indirect exchange, or RKKY interaction, is expressed as
\begin{equation}
	H_{\rm ex} = - \frac{\mathcal{J}^2}{\pi} \, \mathrm{Im} \int_{-\infty}^{\varepsilon_{\rm F}} dE \, \chi(\mathbf{R}, E),
\end{equation}
with the static spin susceptibility (response function) given by
\begin{equation}
	\chi(\mathbf{R}, E) = \mathrm{Tr} \Big[ (\mathbf{M}_1 \cdot \boldsymbol{\tau}) \, G(\mathbf{R}, E) \, (\mathbf{M}_2 \cdot \boldsymbol{\tau}) \, G(-\mathbf{R}, E) \Big],
\end{equation}
where \(G(\mathbf{R}, E)\) is the electronic Green's function in position space derived from the effective Hamiltonian of the system, including spin-orbit and altermagnetic terms, as well as any static lattice renormalizations. They can be obtained from momentum space via a Fourier transform:
\begin{equation}
	G(\mathbf{R}, E) = \frac{1}{4\pi^2} \int d^2 q \, G(\mathbf{q}, E) \, e^{i \mathbf{q} \cdot \mathbf{R}},
\end{equation}where $G(\mathbf{q}, E) = ([E + i\eta]I - H_{\mathbf{q}}^{\rm eff})^{-1}$ with $\eta$ an infinitesimally small parameter. 

The presence of vibrational modes introduces additional complexity by modifying the electronic propagators and the spin-dependent energy landscape. To account for these effects accurately, we perform the momentum-space integrals and energy summation numerically over a dense \(\mathbf{q}\)-grid, ensuring convergence and capturing the full dependence on the Rashba coupling \(\alpha\) and lattice distortions. This allows exploration of regimes in which RSOC is strong and the conventional analytical approximations for RKKY interactions fail.

Decomposition of \(H_{\rm ex}\) reveals contributions of distinct symmetry types. The symmetric components include the Heisenberg term \(\mathbf{M}_1 \cdot \mathbf{M}_2\), Ising-like anisotropies \(\propto M_1^z M_2^z\), and compass-type interactions dependent on crystallographic orientation. In addition, antisymmetric contributions of DM form, \(\mathbf{D}_{\rm DM} \cdot (\mathbf{M}_1 \times \mathbf{M}_2)\), arise due to broken inversion and spin-orbit coupling~\cite{PhysRevB.102.075411,Cheraghchi2021,PhysRevB.110.035307}. These antisymmetric exchange terms are expected to play an important role in stabilizing noncollinear spin configurations, such as spin spirals or skyrmionic textures, in magnetically doped altermagnets. 

\section{Results}\label{s3}\begin{figure}[t]
	\centering
	\includegraphics[width=1\linewidth]{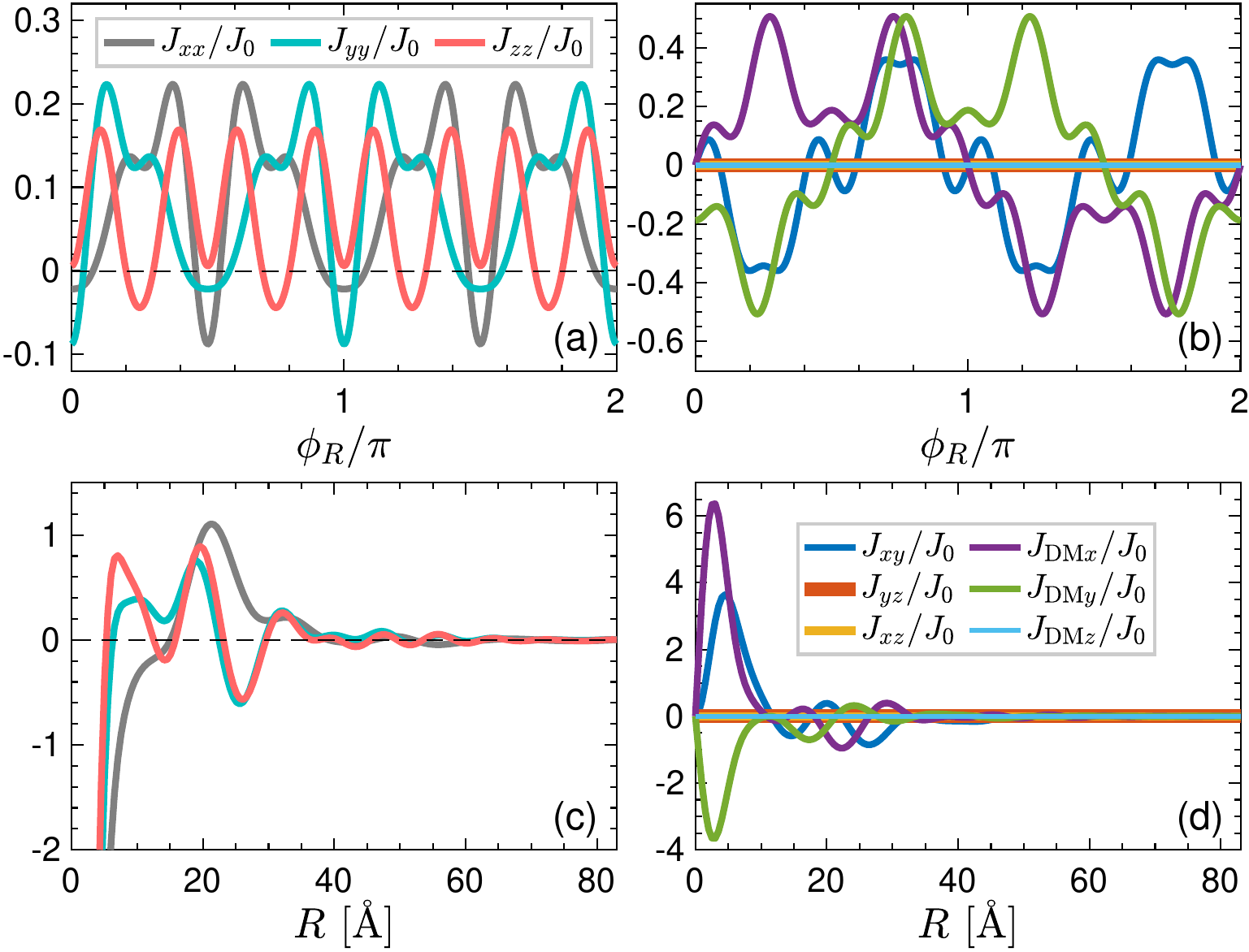}
	\caption{(a,b) Angular dependence of the RKKY tensor components $J_{ij}$ for a pristine 2D Rashba $d_{x^2-y^2}$-wave altermagnet without EPC as a function of the relative orientation angle $\phi_R$ between two magnetic impurities separated by $R = 30$~\AA. 
		(a) Heisenberg and Ising components $J_{xx}$, $J_{yy}$, and $J_{zz}$ exhibit a periodic oscillatory behavior with $\phi_R$ reflecting the underlying spin-split Fermi surface anisotropy. 
		(b) Anisotropic and DM components show finite angular modulations (still periodic) arising from the interplay between altermagnetic spin-splitting and spin-orbit coupling. 
		(c,d) Distance dependence of the corresponding RKKY couplings for fixed $\phi_R = \pi/3$. 
		(c) The diagonal terms display oscillatory decay $1/R^2$ characteristic of indirect exchange interactions mediated by itinerant quasiparticles. 
		(d) The antisymmetric DM components $J_{\mathrm{DM}i}$ emerge from Rashba-induced spin-momentum locking and decay over a similar length scale, exhibiting a phase shift relative to the symmetric exchange parts. The parameters used are $\delta = 0.5$, $\alpha = 0.2$ eV$\cdot$\AA, $\varepsilon_{\rm F} = 0$, $\lambda = 0$, and $\tilde{\lambda}/\lambda = 0$ in (a,b), while (c,d) correspond to the same $\delta$, $\alpha$, and $\varepsilon_F$, but at $\tilde{\lambda}/\lambda = 1/4$, and $\phi_R = \pi/3$.}
	\label{f2}
\end{figure}  
The relative magnitudes and spatial dependence of the RKKY terms are controlled by \(R = |\mathbf{R}|\), the Rashba parameter \(\alpha\), altermagnetic order \(\delta\), the phonon-induced lattice modifications, and doping level $\varepsilon_{\rm F}$, allowing the indirect exchange to be tuned via vibrational or structural engineering. This framework provides a quantitative basis for predicting complex magnetic textures and spin-dependent interactions in altermagnetic and spin-orbit-coupled materials. In what follows, we normalize all components to $J_0 = \mathcal{J}^2 N_q$~($N_q$ being the number of discrete momentum points used in the Brillouin-zone summation) and the sign of the $J_{ij}$ exchange interactions are used to distinguish between ferromagnetic (FM) and antiferromagnetic (AFM) coupling between the two magnetic impurities, corresponding respectively to \( J_{ij} > 0 \) and \( J_{ij} < 0 \). For the DM interactions, we have clockwise~(CW), \( J_{{\rm DM}i} > 0 \), and counterclockwise~(CCW), \( J_{{\rm DM}i} < 0 \), couplings. 

To set the EPC strengths, we employ the 
Lindemann criterion~\cite{Lindemann1910,yarmohammadi2025slowphononcontrolspinedelstein}.
This criterion limits the atomic displacement to roughly 
$10\%$ of the lattice constant, thereby preventing structural instabilities. 
For a typical lattice spacing in $d$-wave altermagnets of 
$a \approx 3\,\text{\AA}$, we select the on-site coupling strength 
$\lambda$ such that the equilibrium lattice displacement satisfies 
$X \lesssim 0.3\,\text{\AA}$. Furthermore, the staggered coupling $\tilde{\lambda}$, which encodes the 
spin-dependent energy splitting arising from the coupling to spin-split 
states, is expected to be weaker than its spin-degenerate counterpart $\lambda$. 
Consistent with this physical hierarchy, we set $\tilde{\lambda}/\lambda = 1/4$ as a 
representative value throughout our calculations, unless stated otherwise.

Figure~\ref{f2} presents the angular and distance dependence of the RKKY tensor components $J_{ij}$ in a 2D Rashba altermagnet, highlighting both the symmetric and antisymmetric exchange channels. For the calculations, the model parameters are chosen as $\delta = 0.5$, $\alpha = 0.2$ eV$\cdot$\AA, $\varepsilon_{\rm F} = 0$, $\lambda = 0$, and $\tilde{\lambda}/\lambda = 0$ for the angular dependence in panels~(a,b), while panels~(c,d) correspond to $\tilde{\lambda}/\lambda = 1/4$ at fixed $\phi_R = \pi/3$.

In Fig.~\ref{f2}(a), the diagonal Heisenberg and Ising components $J_{xx}$, $J_{yy}$, and $J_{zz}$ exhibit oscillatory modulations as a function of $\phi_R$. These different oscillations reflect the anisotropic spin texture of the underlying spin-split Fermi surface, which is a hallmark of Rashba-type coupling in the altermagnetic background. The periodic angular dependence indicates that the indirect magnetic interaction inherits the crystal anisotropy and the spin-momentum locking pattern of the conduction electrons. Notably, the amplitude and phase of $J_{xx}$ and $J_{yy}$ differ slightly, signaling an intrinsic in-plane exchange anisotropy. As the majority of the oscillations occur for $J>0$, we can therefore conclude that the Heisenberg and Ising exchange interactions between the two impurities predominantly favor a FM alignment. Figure~\ref{f2}(b) shows the angular dependence of the anisotropic and DM components, which also display clear oscillatory behavior with $\phi_R$. The finite angular modulations of these off-diagonal terms arise from the interplay between altermagnetic spin-splitting and RSOC. The emergence of DM components ($J_{\mathrm{DM}i}$) is particularly indicative of spin-momentum locking, as their oscillations reflect the chiral symmetry breaking introduced by Rashba interactions. However, unlike the Heisenberg and Ising terms, the \( J_{xy} \) and DM interactions can favor either CW or CCW alignment, depending on the relative orientation of the two impurities.

It is worth noting that the symmetric components \( J_{iz} \) vanish identically as a consequence of the antisymmetric property of the Green’s function products, 
\( G_i(R,E) G_z(-R,E) = -\, G_i(-R,E) G_z(R,E) \). 
An analogous argument applies to the antisymmetric DM component \( J_{{\rm DM}z} \), which also vanishes owing to the odd parity of the out-of-plane Green’s function, 
\( G_z(-R,E) = -\, G_z(R,E) \).

The distance dependence of the same tensor components is also depicted in Figs.~\ref{f2}(c,d) for $\phi_R = \pi/3$. The diagonal components in Fig.~\ref{f2}(c) exhibit damped oscillations with $R$, following the characteristic $1/R^2$ decay of RKKY interactions in two dimensions mediated by itinerant quasiparticles. The oscillatory sign changes correspond to alternating FM and AFM coupling regions, consistent with interference between spin-split Fermi contours. In contrast, Fig.~\ref{f2}(d) illustrates that the DM components decay over a similar length scale but maintain a distinct phase shift relative to the symmetric exchange terms. This phase shift directly reflects the Rashba-induced chirality in the spin-dependent scattering processes. The overall behavior underscores that the Rashba field not only modifies the magnitude of exchange couplings but also introduces a directional asymmetry, enriching the magnetic response landscape of altermagnets.\begin{figure}[t]
			\centering
			\includegraphics[width=1\linewidth]{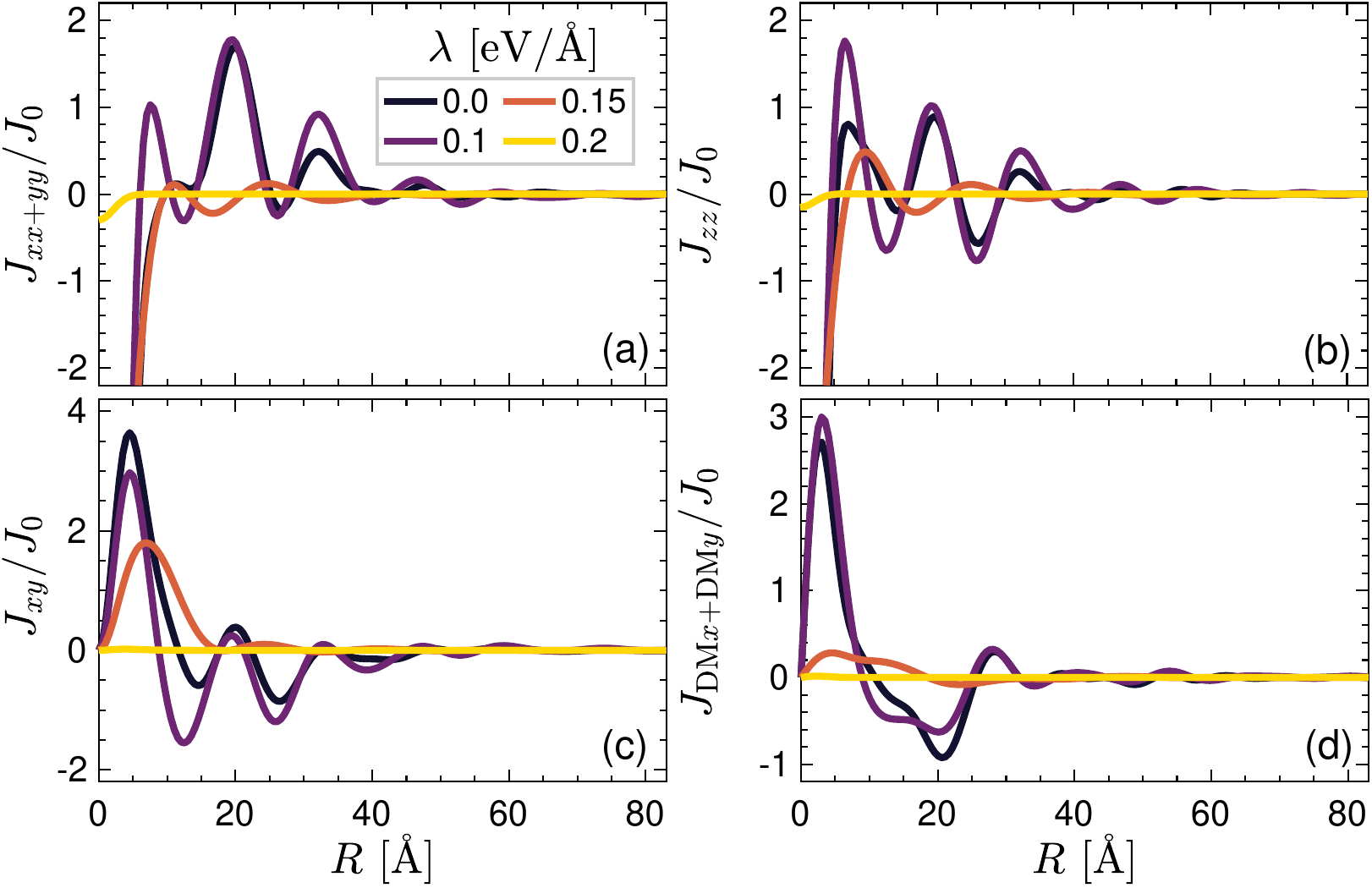}
			\caption{Spatial dependence of RKKY exchange components for several values of EPC \(\lambda\) in a 2D Rashba $d_{x^2-y^2}$-wave altermagnet. Parameters: \(\delta=0.5\), \(\alpha=0.2\) eV$\cdot$\AA, \(\varepsilon_{\rm F}=0\), \(\tilde{\lambda}/\lambda=1/4\), and \(\phi_R=\pi/3\). (a) In-plane symmetric exchange \(J_{xx+yy} = J_{xx}+J_{yy}\), (b) out-of-plane Ising component \(J_{zz}\), (c) off-diagonal symmetric (compass) term \(J_{xy}\), and (d) antisymmetric (DM) combination \(J_{{\rm DM},x+{\rm DM},y} = J_{{\rm DM},x}+J_{{\rm DM},y}\). Increasing EPC ($\lambda$) reduces the long-range coherence and, beyond a moderate coupling strength, suppresses all exchange contributions.}
			\label{f3n}
		\end{figure}  
		
		\subsection{Electron-phonon coupling effects}
Figure~\ref{f3n} shows the spatial dependence of distinct components of the indirect exchange interaction between two localized magnetic moments, obtained from an RKKY calculation based on the Rashba-altermagnet model including EPC. Other model parameters are held fixed at \(\delta = 0.5\), \(\alpha =  0.2\) eV$\cdot$\AA, \(\varepsilon_{\rm F} = 0\), \(\tilde{\lambda}/\lambda = 1/4\), and direction \(\phi_R = \pi/3\). To better distinguish the in-plane and out-of-plane contributions and to avoid redundant features in the plots, we next present the combined in-plane components $J_{xx+yy}$ and $J_{{\rm DM}x+{\rm DM}y}$.

Figure~\ref{f3n}(a) displays the combined in-plane symmetric exchange, which largely captures the isotropic (Heisenberg-like) component of the RKKY tensor in the plane. For \(\lambda=0\) the coupling exhibits the characteristic RKKY oscillations set by the Fermi wavelength and decays algebraically with distance, reflecting coherent, long-ranged conduction-mediated spin correlations. For small EPCs, the responses are close to the pristine case, while as the EPC \(\lambda\) is increased to moderate values, the amplitude of these oscillations is reduced and the long-range tail becomes suppressed. Physically, EPC renormalizes the quasiparticle spectrum, i.e., $\varepsilon_{\mathbf{q},s}^{\rm eff} = \epsilon_{\mathbf{q}} + \lambda X + s \sqrt{ (\gamma_{\mathbf{q}} + \tilde{\lambda} X)^2 + \alpha^2 q^2 }$, and reduces density of itenerant electrons that damp phase-coherent spin-transport, thereby reducing the magnitude and range of the symmetric exchange. Figure~\ref{f3n}(b) shows the out-of-plane (Ising-like) component \(J_{zz}\). Finite \(\lambda\) modifies both amplitude and effective oscillation phase, indicating that vibrational dressing not only attenuates the overall coupling but also alters the effective spin anisotropy through a momentum-dependent renormalization of the altermagnetic splitting, $\gamma_{\mathbf{q}} + \tilde{\lambda} X$.\begin{figure}[t]
	\centering
	\includegraphics[width=1\linewidth]{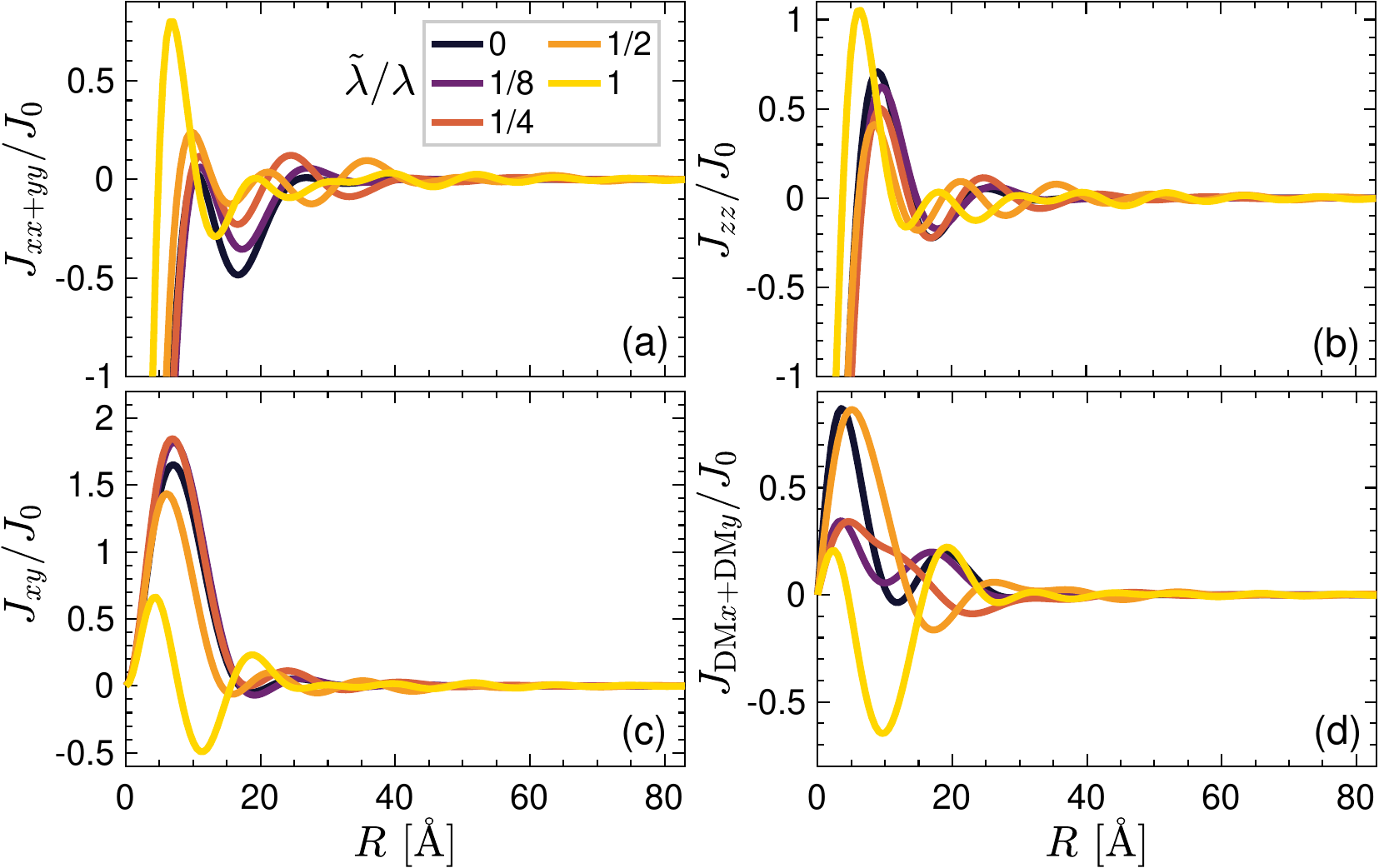}
	\caption{Distance dependence of the phonon-renormalized RKKY interaction components in a 2D Rashba $d_{x^2-y^2}$-wave altermagnet for different $\tilde{\lambda}$, denoting the EPC to the spin-split bands. The parameters are $\delta = 0.5$, $\alpha = 0.2$ eV$\cdot$\AA, $\varepsilon_{\rm{F}} = 0$, $\lambda = 0.15$ eV/\AA, and $\phi_R = \pi/3$. Increasing $\tilde{\lambda}$ both renormalizes amplitudes and shifts the oscillation phases, producing component-specific enhancement, suppression, and even sign changes. }
	\label{f4nn}
\end{figure}  

The off-diagonal symmetric term \(J_{xy}\) is also shown  in Fig.~\ref{f3n}(c). This compass-like interaction vanishes in isotropic media but becomes nonzero when spin-orbit coupling and altermagnetic asymmetry substantially mix spin channels. For instance, in the studied by Amundsen~\textit{et al.}~\cite{PhysRevB.110.054427} and Yarmohammadi~\textit{et al.}~\cite{k3xb-8pts}, it is zero due to weak RSOC. For moderate EPC the amplitude of \(J_{xy}\) is suppressed relative to the \(\lambda=0\) case, signifying that vibrations can not promote spin-channel mixing and thereby decrease anisotropic exchange pathways. Eventually, it destroys the coherence necessary for appreciable long-range anisotropic exchange. Figure~\ref{f3n}~(d) finally presents the antisymmetric in-plane DM contribution. For small \(\lambda\) the DM component shows the expected oscillatory sign changes between CW and CCW phases, while moderate EPC initially suppresses the short-range DM amplitude. When EPC becomes large, however, the DM amplitude is suppressed and the oscillatory structure is washed out, consistent with phonon-induced broadening and dynamical screening of chiral correlations. The adiabatic approximation employed here assumes phonon frequencies much smaller than the electronic bandwidth and Fermi energy. This condition is satisfied for typical optical phonons with frequencies 10--50 meV in candidate altermagnets.

These trends follow directly from how slow (Holstein-type) phonons renormalize quasiparticle properties and hence the $k$-space ingredients of the RKKY kernel. Static EPC (treated here at the Holstein level) renormalizes the quasiparticle residue and effective mass, shifts spectral weight, and introduces additional phase shifts in the scattering amplitudes. Accordingly, the distinct sensitivity of each channel follows from their microscopic scattering processes.
		
Next, Fig.~\ref{f4nn} illustrates various components of the RKKY interaction as a function of the impurity separation \( R \) in a \( d \)-wave altermagnet when the EPC to the spin-split bands is varied. Three distinct phonon-induced effects are exhibited, compared to EPC to the spin-degenerate bands in Fig.~\ref{f3n}, that are component dependent: (i) a strong renormalization of the short-range amplitude, (ii) systematic phase shifts that produce beating patterns at intermediate distances, and (iii) highly non-monotonic and sign-changing behavior of the antisymmetric (DM) and off-diagonal channels. Quantitatively, the in-plane isotropic combination $J_{xx+yy}$ [Fig.~\ref{f4nn}(a)] shows a pronounced increase of the first positive peak as $\tilde{\lambda}/\lambda$ is raised, together with a clear phase advance of subsequent oscillations. The out-of-plane Ising channel $J_{zz}$ [Fig.~\ref{f4nn}(b)] responds oppositely up to $\tilde{\lambda}/\lambda < 1$ but with a slightly different phase shift, so that constructive/destructive interference between curves with different $\tilde\lambda$ produces visible beating in the 10--40 \AA\, range. In contrast, the symmetric off-diagonal $J_{xy}$ [Fig.~\ref{f4nn}(c)] shows a non-monotonic dependence on $\tilde\lambda$: modest phonon coupling ($\tilde\lambda/\lambda\sim 1/8$--$1/4$) slightly enhances the primary peak, whereas strong coupling ($\tilde\lambda/\lambda>1/4$) dramatically suppresses it (peak reduced by a factor $\gtrsim 2$--3) and shifts the phase. Finally, the DM combination [Fig.~\ref{f4nn}(d)] is the most sensitive: with increasing $\tilde\lambda$ the short-range positive peak is reduced and, for the largest coupling, a deep negative trough appears at intermediate $R$, i.e. a clear sign reversal of the DM amplitude.

\subsection{Altermagnetic strength effects}

Figure~\ref{f4n} shows the evolution of the different components of the exchange interaction as a function of the EPC strength $\lambda$ for several values of the altermagnetic order parameter $\delta$. A careful inspection reveals that the amplitude and sign respond differently to increasing $\delta$, reflecting the nontrivial interplay between electron-phonon interactions and altermagnetic order. For $\delta = 0$ (Rashba non-altermagnet phase), the system corresponds to a Rashba spin-degenerate background with EPC as the only interaction. In this case, all exchange components display smooth and relatively low-amplitude oscillations as a function of $\lambda$, indicative of well-defined phonon-mediated indirect exchange processes in a spin-symmetric band structure. As $\delta$ increases to $0.4$ and further to $0.8$, the oscillatory pattern, mainly responsible for sign of interactions, becomes significantly more structured. Both $J_{xx+yy}$ and $J_{zz}$ exhibit enhanced amplitude and sharper features around intermediate $\lambda$ values ($0.05$--$0.15$~eV/\AA), showing that altermagnetic order strongly modulates the strength and phase of phonon-mediated exchange.\begin{figure}[t]
	\centering
	\includegraphics[width=1\linewidth]{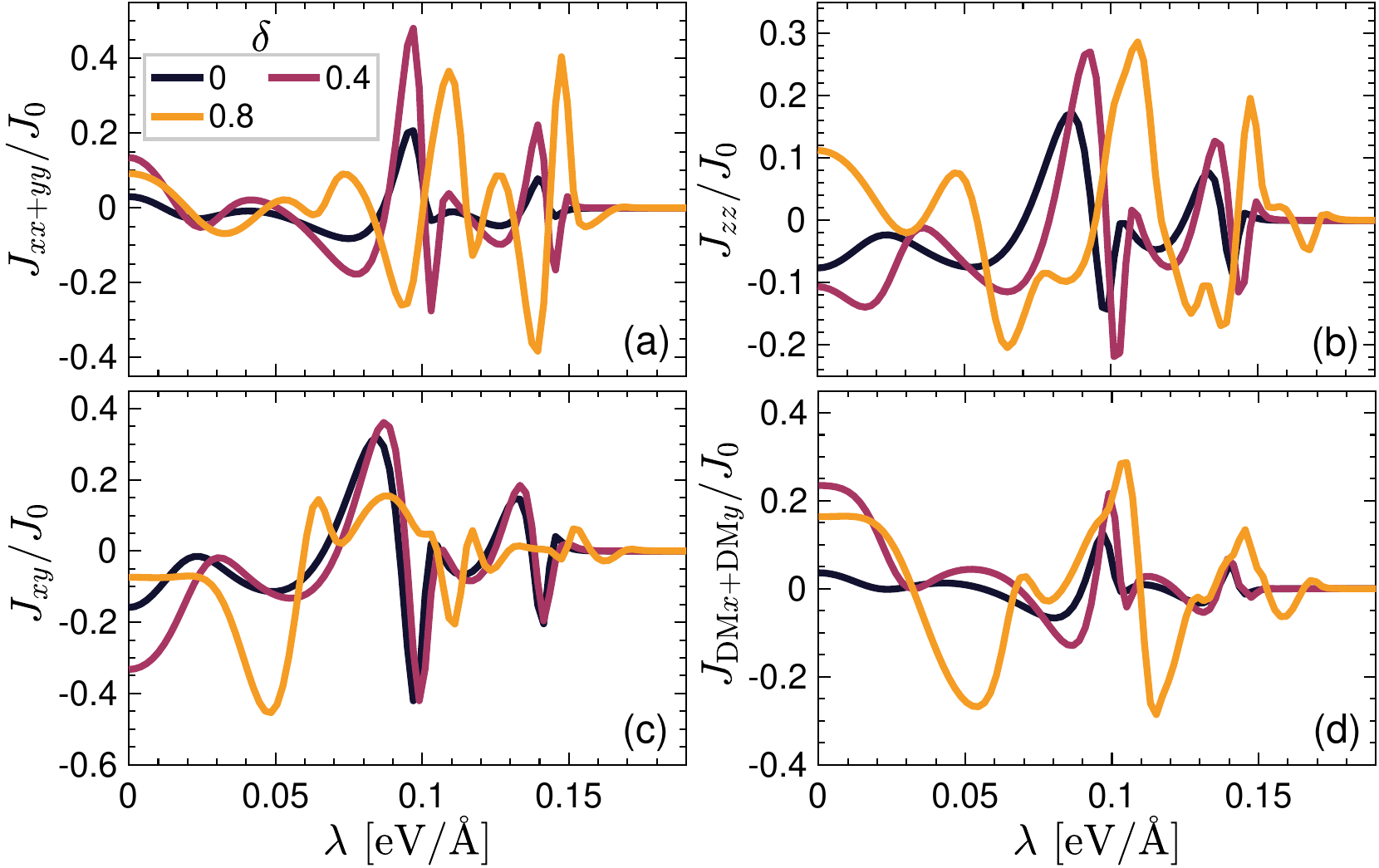}
	\caption{Evolution of the different components of the RKKY exchange interaction as a function of the EPC strength $\lambda$ for different values of the altermagnetic order parameter $\delta$. Increasing $\delta$ leads to a pronounced enhancement of both the amplitude and complexity of the oscillations in exchange channels, reflecting the strong modulation of phonon-mediated interactions by the altermagnetic order. Parameters: $\alpha = 0.2$~eV$\cdot$\AA, $\varepsilon_{\rm F}=0$, $\tilde{\lambda}/\lambda = 1/4$, $\phi_R = \pi/3$, and $R = 30$~\AA.}
	\label{f4n}
\end{figure}  

Interestingly, the antisymmetric components $J_{xy}$ and $J_{\rm DMx+DM y}$ display a more robust and less distorted oscillatory structure as $\delta$ increases, although their overall magnitude does increase. This indicates that while altermagnetic order enhances the strength of the DM-like contributions, it does not drastically modify their underlying oscillatory character. In contrast, the symmetric exchange channels are highly sensitive to $\delta$, exhibiting not only amplitude growth but also clear phase shifts and more irregular oscillations. This suggests that the altermagnetic background acts as an effective internal field that distorts the phonon-mediated exchange pathways.\begin{figure}[t]
			\centering
			\includegraphics[width=1\linewidth]{Fig6_new.pdf}
			\caption{Dependence of the RKKY interaction components on the EPC strength \(\lambda\) for different RSOC strengths \(\alpha\). Here, \(\delta = 0.5\), \(\varepsilon_F = 0\), \(\tilde{\lambda}/\lambda = 1/4\), \(\phi_R = \pi/3\), and \(R = 30~\text{\AA}\). For vanishing RSOC (\(\alpha = 0\)), the RKKY interactions are predominantly isotropic, with the in-plane Heisenberg and out-of-plane Ising terms showing small amplitude oscillations as a function of \(\lambda\). Increasing \(\alpha\) enhances anisotropic effects, leading to larger amplitudes and pronounced oscillations in all components, especially in the off-diagonal \(J_{xy}\) and DM terms, indicating that RSOC significantly modulates the spin texture of the indirect exchange.}
			\label{f5n}
		\end{figure}  

\subsection{Rashba spin-orbit coupling effects}

The behavior of the RKKY interaction components as a function of the EPC strength \(\lambda\) for various RSOC strenghts are shown in Fig.~\ref{f5n}. For vanishing RSOC, the exchange interaction remains nearly isotropic, and both the in-plane Heisenberg \((J_{xx}+J_{yy})\) and out-of-plane Ising \((J_{zz})\) components exhibit weak, smooth oscillations with \(\lambda\). 
In this limit, the spin degeneracy of the electronic bands ensures that the RKKY coupling is primarily governed by the scalar part of the electronic susceptibility, resulting in minimal anisotropy and negligible DM contributions. Once RSOC is introduced, the spin degeneracy of the mediating electrons is lifted, leading to spin-momentum locking and a substantial modification of all RKKY components. The finite \(\alpha\) generates anisotropic exchange interactions and enhances the amplitude of the oscillations with \(\lambda\), as observed in all panels of Fig.~\ref{f5n}. Notably, the in-plane off-diagonal term \(J_{xy}\) and the DM components acquire pronounced oscillations and even sign reversals, demonstrating that the RSOC induces an effective chiral exchange between impurities. 
This behavior originates from interference between spin-split Fermi contours, which produces additional oscillatory length scales in the RKKY range function. Thus, the combined effect of RSOC and EPC provides an efficient handle to control magnetic correlations between impurities. 
By adjusting \(\alpha\) and \(\lambda\), one can continuously tune the system between FM/CW and AFM/CCW coupling regimes, or even realize noncollinear spin textures stabilized by the DM interaction.\begin{figure}[t]
	\centering
	\includegraphics[width=1\linewidth]{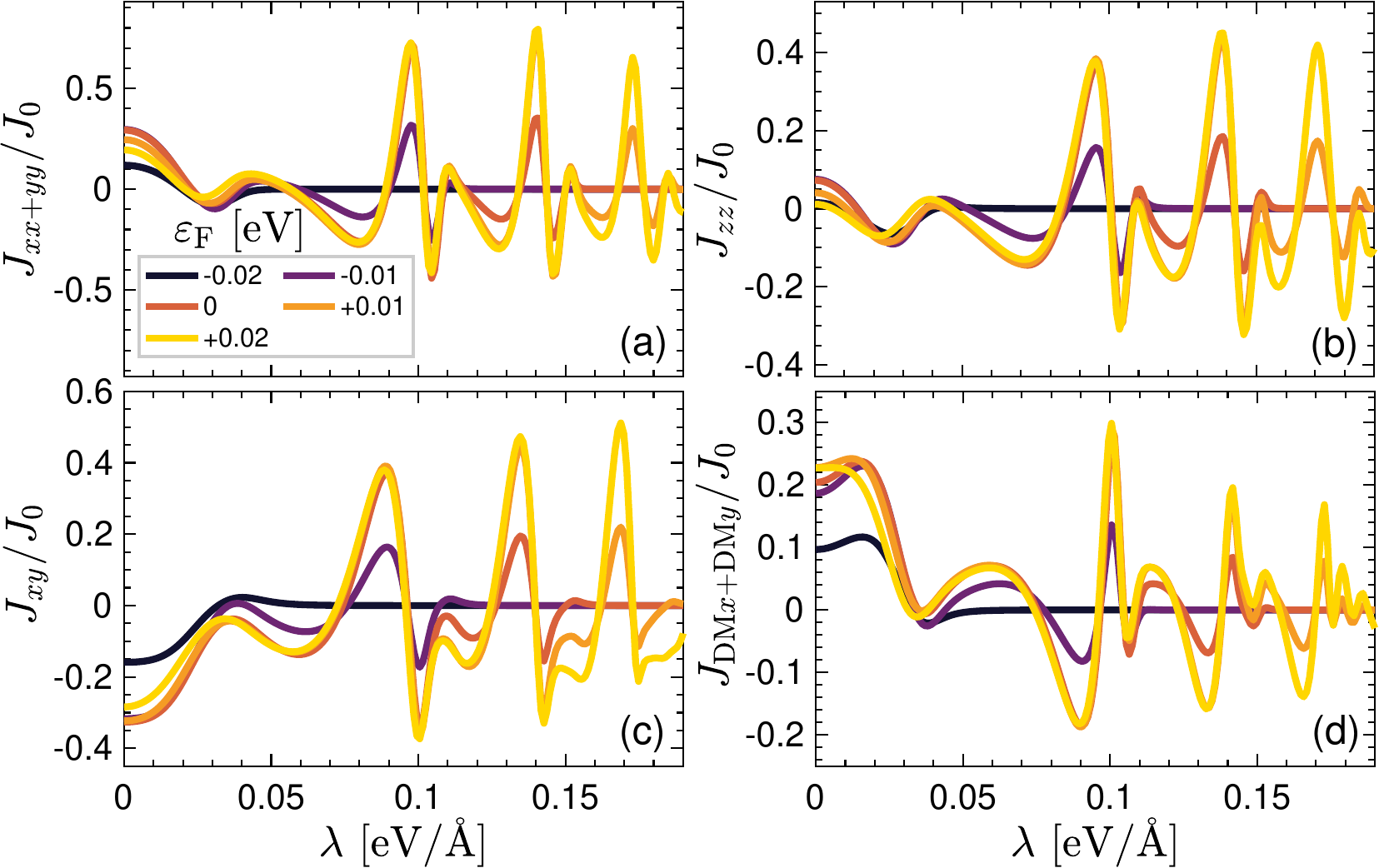}
	\caption{Dependence of the normalized RKKY interaction components on the EPC strength \(\lambda\) for several values of the chemical potential (Fermi energy) \(\varepsilon_{\rm F}\). Parameters are \(\delta = 0.5\), \(\alpha = 0.2\) eV$\cdot$\AA, \(\tilde{\lambda}/\lambda = 1/4\), \(\phi_R = \pi/3\), and \(R = 30~\text{\AA}\).
		Increasing \(\varepsilon_{\rm F}\) modifies the oscillation amplitude and period of all RKKY components by altering the Fermi wave vector of the spin-split Rashba bands. 
		Positive (negative) \(\varepsilon_{\rm F}\) shifts, corresponding to electron (hole) doping, induce phase displacements of opposite sign, reflecting the asymmetric contribution of the upper and lower Rashba branches to the exchange.}
	\label{f6}
\end{figure}  
		
		\subsection{Doping effects}
		
Finally, Fig.~\ref{f6} illustrates the influence of the Fermi energy \(\varepsilon_{\rm F}\) on the RKKY interaction components for a fixed RSOC and EPCs strength. 
At zero temperature, \(\varepsilon_{\rm F}\) represents the chemical potential controlling the occupation of the spin-split bands, and its variation effectively changes the phase accumulation of the itinerant carriers mediating the magnetic coupling. 
As a result, the overall magnitude and phase of the RKKY oscillations become highly sensitive to small shifts in \(\varepsilon_{\rm F}\). When \(\varepsilon_{\rm F}=0\), corresponding to the charge-neutral condition, the RKKY components exhibit moderately damped oscillations as a function of \(\lambda\). 
Introducing a finite \(\varepsilon_{\rm F}\) enhances the oscillation amplitude and modifies the period of the exchange components. Positive and negative shifts of \(\varepsilon_{\rm F}\) yield opposite phase displacements, corresponding to electron and hole doping, reflecting the asymmetric contribution of the upper and lower Rashba branches to the exchange pathways. 
This asymmetry is most evident at larger EPC strengths, where the EPC modulates the effective band curvature and further enhances interference effects between multiple scattering channels.

The evolution with \(\varepsilon_{\rm F}\) thus highlights the tunability of the indirect magnetic interaction via carrier density. 
By doping it away from charge neutrality, one can reversibly control both the magnitude and the sign of the exchange couplings, enabling electrical control of the transition between FM/CW, AFM/CCW, and chiral spin-aligned configurations. The combined sensitivity to \(\varepsilon_{\rm F}\), \(\alpha\), and \(\lambda\) underscores the cooperative nature of spin-orbit and electron-phonon mechanisms in engineering carrier-mediated magnetism in spin-split systems.

		\section{Conclusions and outlook}\label{s4}
		In this work, we have systematically elucidated the profound influence of electron-phonon coupling (EPC) on the noncollinear Ruderman-Kittel-Kasuya-Yosida (RKKY) interaction in 2D Rashba $d$-wave altermagnets, bridging a critical gap in understanding how lattice vibrations mediate spin-dependent impurity exchanges under arbitrary Rashba spin-orbit coupling (RSOC) regimes. By extending a continuum model to incorporate spin-selective Holstein-type EPC and treating lattice distortions adiabatically, we derived a renormalized electronic spectrum that self-consistently captures phonon feedback on quasiparticle dispersions and spin polarizations. Numerical evaluation of the RKKY tensor reveals a rich variety of tunable exchange anisotropies. While the symmetric Heisenberg and Ising components display oscillations with phonon-suppressed long-range tails, the antisymmetric Dzyaloshinskii–Moriya (DM) and off-diagonal terms exhibit pronounced phase shifts, non-monotonic amplitude enhancements, and sign inversions — signatures of interference between spin-momentum-locked Fermi contours dressed by lattice vibrations.
		
		Key insights highlight the differential sensitivity of exchange channels to EPC strengths: charge-coupled phonons generally suppress coherence, whereas spin-coupled phonons induce beating patterns and chiral reversals. These behaviors synergize with altermagnetic order, RSOC, and the chemical potential to tune the oscillatory complexity and allow selective stabilization of FM/AFM or CW/CCW configurations. These results not only affirm the inadequacy of weak-RSOC approximations for capturing full anisotropy but also demonstrate phonons as a low-energy, gate-tunable lever for sculpting magnetic interactions, surpassing optical or field-based methods in static precision. Ultimately, our framework establishes phononic engineering as a cornerstone for realizing emergent spin orders in altermagnets, fostering advances in topological spintronics and correlated quantum materials.
		
		Future extensions of this study could incorporate dynamical EPC beyond the static-Holstein approximation, such as via Migdal-Eliashberg theory, to explore frequency-dependent RKKY responses under THz phonon driving, potentially unveiling Floquet-engineered chiralities or polaronic enhancements in DM terms. Moreover, embedding our model in multilayer geometries with interfacial RSOC gradients could reveal proximity-induced gradients in RKKY anisotropy, paving the way for designer skyrmion lattices or magnonic crystals. On the theoretical front, extending to higher-order exchanges or finite-temperature effects may uncover EPC-mediated superconducting pairings intertwined with noncollinear magnetism. Collectively, these directions position phononic control of altermagnetic RKKY as a linchpin for next-generation spin-valve devices.
		
		\section*{Acknowledgments}
		This work was supported by Hue University under the Core Research Program, Grant No. NCTB.DHH.2025.17.
		\section*{Conflicts of interest}
			There are no conflicts to declare.
	\section*{Data availability statement}
	The data that support the findings of this study are available upon reasonable request from the authors.
	}
		\bibliography{bib.bib}

\begin{thebibliography}{59}%
\makeatletter
\providecommand \@ifxundefined [1]{%
 \@ifx{#1\undefined}
}%
\providecommand \@ifnum [1]{%
 \ifnum #1\expandafter \@firstoftwo
 \else \expandafter \@secondoftwo
 \fi
}%
\providecommand \@ifx [1]{%
 \ifx #1\expandafter \@firstoftwo
 \else \expandafter \@secondoftwo
 \fi
}%
\providecommand \natexlab [1]{#1}%
\providecommand \enquote  [1]{``#1''}%
\providecommand \bibnamefont  [1]{#1}%
\providecommand \bibfnamefont [1]{#1}%
\providecommand \citenamefont [1]{#1}%
\providecommand \href@noop [0]{\@secondoftwo}%
\providecommand \href [0]{\begingroup \@sanitize@url \@href}%
\providecommand \@href[1]{\@@startlink{#1}\@@href}%
\providecommand \@@href[1]{\endgroup#1\@@endlink}%
\providecommand \@sanitize@url [0]{\catcode `\\12\catcode `\$12\catcode
  `\&12\catcode `\#12\catcode `\^12\catcode `\_12\catcode `\%12\relax}%
\providecommand \@@startlink[1]{}%
\providecommand \@@endlink[0]{}%
\providecommand \url  [0]{\begingroup\@sanitize@url \@url }%
\providecommand \@url [1]{\endgroup\@href {#1}{\urlprefix }}%
\providecommand \urlprefix  [0]{URL }%
\providecommand \Eprint [0]{\href }%
\providecommand \doibase [0]{https://doi.org/}%
\providecommand \selectlanguage [0]{\@gobble}%
\providecommand \bibinfo  [0]{\@secondoftwo}%
\providecommand \bibfield  [0]{\@secondoftwo}%
\providecommand \translation [1]{[#1]}%
\providecommand \BibitemOpen [0]{}%
\providecommand \bibitemStop [0]{}%
\providecommand \bibitemNoStop [0]{.\EOS\space}%
\providecommand \EOS [0]{\spacefactor3000\relax}%
\providecommand \BibitemShut  [1]{\csname bibitem#1\endcsname}%
\let\auto@bib@innerbib\@empty
\bibitem [{\citenamefont {Fert}\ \emph {et~al.}(2017)\citenamefont {Fert},
  \citenamefont {Reyren},\ and\ \citenamefont {Cros}}]{Fert2017}%
  \BibitemOpen
  \bibfield  {author} {\bibinfo {author} {\bibfnamefont {A.}~\bibnamefont
  {Fert}}, \bibinfo {author} {\bibfnamefont {N.}~\bibnamefont {Reyren}},\ and\
  \bibinfo {author} {\bibfnamefont {V.}~\bibnamefont {Cros}},\ }\bibfield
  {title} {\bibinfo {title} {Magnetic skyrmions: advances in physics and
  potential applications},\ }\href {https://doi.org/10.1038/natrevmats.2017.31}
  {\bibfield  {journal} {\bibinfo  {journal} {Nature Reviews Materials}\
  }\textbf {\bibinfo {volume} {2}},\ \bibinfo {pages} {17031} (\bibinfo {year}
  {2017})}\BibitemShut {NoStop}%
\bibitem [{\citenamefont {Jungwirth}\ \emph {et~al.}(2016)\citenamefont
  {Jungwirth}, \citenamefont {Marti}, \citenamefont {Wadley},\ and\
  \citenamefont {Wunderlich}}]{Jungwirth2016}%
  \BibitemOpen
  \bibfield  {author} {\bibinfo {author} {\bibfnamefont {T.}~\bibnamefont
  {Jungwirth}}, \bibinfo {author} {\bibfnamefont {X.}~\bibnamefont {Marti}},
  \bibinfo {author} {\bibfnamefont {P.}~\bibnamefont {Wadley}},\ and\ \bibinfo
  {author} {\bibfnamefont {J.}~\bibnamefont {Wunderlich}},\ }\bibfield  {title}
  {\bibinfo {title} {Antiferromagnetic spintronics},\ }\href
  {https://doi.org/10.1038/nnano.2016.18} {\bibfield  {journal} {\bibinfo
  {journal} {Nature Nanotechnology}\ }\textbf {\bibinfo {volume} {11}},\
  \bibinfo {pages} {231} (\bibinfo {year} {2016})}\BibitemShut {NoStop}%
\bibitem [{\citenamefont {\ifmmode \check{Z}\else
  \v{Z}\fi{}uti\ifmmode~\acute{c}\else \'{c}\fi{}}\ \emph
  {et~al.}(2004)\citenamefont {\ifmmode \check{Z}\else
  \v{Z}\fi{}uti\ifmmode~\acute{c}\else \'{c}\fi{}}, \citenamefont {Fabian},\
  and\ \citenamefont {Das~Sarma}}]{RevModPhys.76.323}%
  \BibitemOpen
  \bibfield  {author} {\bibinfo {author} {\bibfnamefont {I.}~\bibnamefont
  {\ifmmode \check{Z}\else \v{Z}\fi{}uti\ifmmode~\acute{c}\else \'{c}\fi{}}},
  \bibinfo {author} {\bibfnamefont {J.}~\bibnamefont {Fabian}},\ and\ \bibinfo
  {author} {\bibfnamefont {S.}~\bibnamefont {Das~Sarma}},\ }\bibfield  {title}
  {\bibinfo {title} {Spintronics: Fundamentals and applications},\ }\href
  {https://doi.org/10.1103/RevModPhys.76.323} {\bibfield  {journal} {\bibinfo
  {journal} {Rev. Mod. Phys.}\ }\textbf {\bibinfo {volume} {76}},\ \bibinfo
  {pages} {323} (\bibinfo {year} {2004})}\BibitemShut {NoStop}%
\bibitem [{\citenamefont {Baltz}\ \emph {et~al.}(2018)\citenamefont {Baltz},
  \citenamefont {Manchon}, \citenamefont {Tsoi}, \citenamefont {Moriyama},
  \citenamefont {Ono},\ and\ \citenamefont
  {Tserkovnyak}}]{RevModPhys.90.015005}%
  \BibitemOpen
  \bibfield  {author} {\bibinfo {author} {\bibfnamefont {V.}~\bibnamefont
  {Baltz}}, \bibinfo {author} {\bibfnamefont {A.}~\bibnamefont {Manchon}},
  \bibinfo {author} {\bibfnamefont {M.}~\bibnamefont {Tsoi}}, \bibinfo {author}
  {\bibfnamefont {T.}~\bibnamefont {Moriyama}}, \bibinfo {author}
  {\bibfnamefont {T.}~\bibnamefont {Ono}},\ and\ \bibinfo {author}
  {\bibfnamefont {Y.}~\bibnamefont {Tserkovnyak}},\ }\bibfield  {title}
  {\bibinfo {title} {Antiferromagnetic spintronics},\ }\href
  {https://doi.org/10.1103/RevModPhys.90.015005} {\bibfield  {journal}
  {\bibinfo  {journal} {Rev. Mod. Phys.}\ }\textbf {\bibinfo {volume} {90}},\
  \bibinfo {pages} {015005} (\bibinfo {year} {2018})}\BibitemShut {NoStop}%
\bibitem [{\citenamefont {\ifmmode~\check{S}\else \v{S}\fi{}mejkal}\ \emph
  {et~al.}(2022{\natexlab{a}})\citenamefont {\ifmmode~\check{S}\else
  \v{S}\fi{}mejkal}, \citenamefont {Sinova},\ and\ \citenamefont
  {Jungwirth}}]{PhysRevX.12.040501}%
  \BibitemOpen
  \bibfield  {author} {\bibinfo {author} {\bibfnamefont {L.}~\bibnamefont
  {\ifmmode~\check{S}\else \v{S}\fi{}mejkal}}, \bibinfo {author} {\bibfnamefont
  {J.}~\bibnamefont {Sinova}},\ and\ \bibinfo {author} {\bibfnamefont
  {T.}~\bibnamefont {Jungwirth}},\ }\bibfield  {title} {\bibinfo {title}
  {Emerging research landscape of altermagnetism},\ }\href
  {https://doi.org/10.1103/PhysRevX.12.040501} {\bibfield  {journal} {\bibinfo
  {journal} {Phys. Rev. X}\ }\textbf {\bibinfo {volume} {12}},\ \bibinfo
  {pages} {040501} (\bibinfo {year} {2022}{\natexlab{a}})}\BibitemShut
  {NoStop}%
\bibitem [{\citenamefont {Mazin}(2022)}]{PhysRevX.12.040002}%
  \BibitemOpen
  \bibfield  {author} {\bibinfo {author} {\bibfnamefont {I.}~\bibnamefont
  {Mazin}} (\bibinfo {collaboration} {The PRX Editors}),\ }\bibfield  {title}
  {\bibinfo {title} {Editorial: Altermagnetism---a new punch line of
  fundamental magnetism},\ }\href {https://doi.org/10.1103/PhysRevX.12.040002}
  {\bibfield  {journal} {\bibinfo  {journal} {Phys. Rev. X}\ }\textbf {\bibinfo
  {volume} {12}},\ \bibinfo {pages} {040002} (\bibinfo {year}
  {2022})}\BibitemShut {NoStop}%
\bibitem [{\citenamefont {\ifmmode~\check{S}\else \v{S}\fi{}mejkal}\ \emph
  {et~al.}(2022{\natexlab{b}})\citenamefont {\ifmmode~\check{S}\else
  \v{S}\fi{}mejkal}, \citenamefont {Sinova},\ and\ \citenamefont
  {Jungwirth}}]{PhysRevX.12.031042}%
  \BibitemOpen
  \bibfield  {author} {\bibinfo {author} {\bibfnamefont {L.}~\bibnamefont
  {\ifmmode~\check{S}\else \v{S}\fi{}mejkal}}, \bibinfo {author} {\bibfnamefont
  {J.}~\bibnamefont {Sinova}},\ and\ \bibinfo {author} {\bibfnamefont
  {T.}~\bibnamefont {Jungwirth}},\ }\bibfield  {title} {\bibinfo {title}
  {Beyond conventional ferromagnetism and antiferromagnetism: A phase with
  nonrelativistic spin and crystal rotation symmetry},\ }\href
  {https://doi.org/10.1103/PhysRevX.12.031042} {\bibfield  {journal} {\bibinfo
  {journal} {Phys. Rev. X}\ }\textbf {\bibinfo {volume} {12}},\ \bibinfo
  {pages} {031042} (\bibinfo {year} {2022}{\natexlab{b}})}\BibitemShut
  {NoStop}%
\bibitem [{\citenamefont {Hayami}\ \emph {et~al.}(2019)\citenamefont {Hayami},
  \citenamefont {Yanagi},\ and\ \citenamefont
  {Kusunose}}]{doi:10.7566/JPSJ.88.123702}%
  \BibitemOpen
  \bibfield  {author} {\bibinfo {author} {\bibfnamefont {S.}~\bibnamefont
  {Hayami}}, \bibinfo {author} {\bibfnamefont {Y.}~\bibnamefont {Yanagi}},\
  and\ \bibinfo {author} {\bibfnamefont {H.}~\bibnamefont {Kusunose}},\
  }\bibfield  {title} {\bibinfo {title} {Momentum-dependent spin splitting by
  collinear antiferromagnetic ordering},\ }\href
  {https://doi.org/10.7566/JPSJ.88.123702} {\bibfield  {journal} {\bibinfo
  {journal} {Journal of the Physical Society of Japan}\ }\textbf {\bibinfo
  {volume} {88}},\ \bibinfo {pages} {123702} (\bibinfo {year}
  {2019})}\BibitemShut {NoStop}%
\bibitem [{\citenamefont {\ifmmode~\check{S}\else \v{S}\fi{}mejkal}\ \emph
  {et~al.}(2022{\natexlab{c}})\citenamefont {\ifmmode~\check{S}\else
  \v{S}\fi{}mejkal}, \citenamefont {Hellenes}, \citenamefont
  {Gonz\'alez-Hern\'andez}, \citenamefont {Sinova},\ and\ \citenamefont
  {Jungwirth}}]{PhysRevX.12.011028}%
  \BibitemOpen
  \bibfield  {author} {\bibinfo {author} {\bibfnamefont {L.}~\bibnamefont
  {\ifmmode~\check{S}\else \v{S}\fi{}mejkal}}, \bibinfo {author} {\bibfnamefont
  {A.~B.}\ \bibnamefont {Hellenes}}, \bibinfo {author} {\bibfnamefont
  {R.}~\bibnamefont {Gonz\'alez-Hern\'andez}}, \bibinfo {author} {\bibfnamefont
  {J.}~\bibnamefont {Sinova}},\ and\ \bibinfo {author} {\bibfnamefont
  {T.}~\bibnamefont {Jungwirth}},\ }\bibfield  {title} {\bibinfo {title} {Giant
  and tunneling magnetoresistance in unconventional collinear antiferromagnets
  with nonrelativistic spin-momentum coupling},\ }\href
  {https://doi.org/10.1103/PhysRevX.12.011028} {\bibfield  {journal} {\bibinfo
  {journal} {Phys. Rev. X}\ }\textbf {\bibinfo {volume} {12}},\ \bibinfo
  {pages} {011028} (\bibinfo {year} {2022}{\natexlab{c}})}\BibitemShut
  {NoStop}%
\bibitem [{\citenamefont {Yuan}\ \emph {et~al.}(2020)\citenamefont {Yuan},
  \citenamefont {Wang}, \citenamefont {Luo}, \citenamefont {{R}ashba},\ and\
  \citenamefont {Zunger}}]{PhysRevB.102.014422}%
  \BibitemOpen
  \bibfield  {author} {\bibinfo {author} {\bibfnamefont {L.-D.}\ \bibnamefont
  {Yuan}}, \bibinfo {author} {\bibfnamefont {Z.}~\bibnamefont {Wang}}, \bibinfo
  {author} {\bibfnamefont {J.-W.}\ \bibnamefont {Luo}}, \bibinfo {author}
  {\bibfnamefont {E.~I.}\ \bibnamefont {{R}ashba}},\ and\ \bibinfo {author}
  {\bibfnamefont {A.}~\bibnamefont {Zunger}},\ }\bibfield  {title} {\bibinfo
  {title} {Giant momentum-dependent spin splitting in centrosymmetric low-$z$
  antiferromagnets},\ }\href {https://doi.org/10.1103/PhysRevB.102.014422}
  {\bibfield  {journal} {\bibinfo  {journal} {Phys. Rev. B}\ }\textbf {\bibinfo
  {volume} {102}},\ \bibinfo {pages} {014422} (\bibinfo {year}
  {2020})}\BibitemShut {NoStop}%
\bibitem [{\citenamefont {Ahn}\ \emph {et~al.}(2019)\citenamefont {Ahn},
  \citenamefont {Hariki}, \citenamefont {Lee},\ and\ \citenamefont
  {Kune\ifmmode~\check{s}\else \v{s}\fi{}}}]{PhysRevB.99.184432}%
  \BibitemOpen
  \bibfield  {author} {\bibinfo {author} {\bibfnamefont {K.-H.}\ \bibnamefont
  {Ahn}}, \bibinfo {author} {\bibfnamefont {A.}~\bibnamefont {Hariki}},
  \bibinfo {author} {\bibfnamefont {K.-W.}\ \bibnamefont {Lee}},\ and\ \bibinfo
  {author} {\bibfnamefont {J.}~\bibnamefont {Kune\ifmmode~\check{s}\else
  \v{s}\fi{}}},\ }\bibfield  {title} {\bibinfo {title} {Antiferromagnetism in
  {R}u{O}$_2$ as $d$-wave pomeranchuk instability},\ }\href
  {https://doi.org/10.1103/PhysRevB.99.184432} {\bibfield  {journal} {\bibinfo
  {journal} {Phys. Rev. B}\ }\textbf {\bibinfo {volume} {99}},\ \bibinfo
  {pages} {184432} (\bibinfo {year} {2019})}\BibitemShut {NoStop}%
\bibitem [{\citenamefont {Yuan}\ \emph {et~al.}(2021)\citenamefont {Yuan},
  \citenamefont {Wang}, \citenamefont {Luo},\ and\ \citenamefont
  {Zunger}}]{PhysRevMaterials.5.014409}%
  \BibitemOpen
  \bibfield  {author} {\bibinfo {author} {\bibfnamefont {L.-D.}\ \bibnamefont
  {Yuan}}, \bibinfo {author} {\bibfnamefont {Z.}~\bibnamefont {Wang}}, \bibinfo
  {author} {\bibfnamefont {J.-W.}\ \bibnamefont {Luo}},\ and\ \bibinfo {author}
  {\bibfnamefont {A.}~\bibnamefont {Zunger}},\ }\bibfield  {title} {\bibinfo
  {title} {Prediction of low-{Z} collinear and noncollinear antiferromagnetic
  compounds having momentum-dependent spin splitting even without spin-orbit
  coupling},\ }\href {https://doi.org/10.1103/PhysRevMaterials.5.014409}
  {\bibfield  {journal} {\bibinfo  {journal} {Phys. Rev. Mater.}\ }\textbf
  {\bibinfo {volume} {5}},\ \bibinfo {pages} {014409} (\bibinfo {year}
  {2021})}\BibitemShut {NoStop}%
\bibitem [{\citenamefont {He}\ \emph {et~al.}(2025)\citenamefont {He},
  \citenamefont {Zhao}, \citenamefont {Luo},\ and\ \citenamefont
  {Hu}}]{he2025altermagnetismttprimedeltafermihubbardmodel}%
  \BibitemOpen
  \bibfield  {author} {\bibinfo {author} {\bibfnamefont {S.}~\bibnamefont
  {He}}, \bibinfo {author} {\bibfnamefont {J.}~\bibnamefont {Zhao}}, \bibinfo
  {author} {\bibfnamefont {H.-G.}\ \bibnamefont {Luo}},\ and\ \bibinfo {author}
  {\bibfnamefont {S.}~\bibnamefont {Hu}},\ }\bibfield  {title} {\bibinfo
  {title} {Altermagnetism and beyond in the
  $t\text{\ensuremath{-}}{t}^{\ensuremath{'}}\text{\ensuremath{-}}\ensuremath{\delta}$
  {F}ermi-{H}ubbard model},\ }\href {https://doi.org/10.1103/4mv8-tb66}
  {\bibfield  {journal} {\bibinfo  {journal} {Phys. Rev. B}\ }\textbf {\bibinfo
  {volume} {112}},\ \bibinfo {pages} {035108} (\bibinfo {year}
  {2025})}\BibitemShut {NoStop}%
\bibitem [{\citenamefont {Yarmohammadi}\ \emph
  {et~al.}(2026{\natexlab{a}})\citenamefont {Yarmohammadi}, \citenamefont
  {\ifmmode~\check{S}\else \v{S}\fi{}mejkal},\ and\ \citenamefont
  {Freericks}}]{7bss-9yxb}%
  \BibitemOpen
  \bibfield  {author} {\bibinfo {author} {\bibfnamefont {M.}~\bibnamefont
  {Yarmohammadi}}, \bibinfo {author} {\bibfnamefont {L.}~\bibnamefont
  {\ifmmode~\check{S}\else \v{S}\fi{}mejkal}},\ and\ \bibinfo {author}
  {\bibfnamefont {J.~K.}\ \bibnamefont {Freericks}},\ }\bibfield  {title}
  {\bibinfo {title} {Cavity-induced coherent magnetization and polaritons in
  altermagnets},\ }\href {https://doi.org/10.1103/7bss-9yxb} {\bibfield
  {journal} {\bibinfo  {journal} {Phys. Rev. Lett.}\ }\textbf {\bibinfo
  {volume} {136}},\ \bibinfo {pages} {146904} (\bibinfo {year}
  {2026}{\natexlab{a}})}\BibitemShut {NoStop}%
\bibitem [{\citenamefont {Bai}\ \emph {et~al.}(2023)\citenamefont {Bai},
  \citenamefont {Zhang}, \citenamefont {Zhou}, \citenamefont {Chen},
  \citenamefont {Wan}, \citenamefont {Han}, \citenamefont {Zhu}, \citenamefont
  {Liang}, \citenamefont {Su}, \citenamefont {Han}, \citenamefont {Pan},\ and\
  \citenamefont {Song}}]{PhysRevLett.130.216701}%
  \BibitemOpen
  \bibfield  {author} {\bibinfo {author} {\bibfnamefont {H.}~\bibnamefont
  {Bai}}, \bibinfo {author} {\bibfnamefont {Y.~C.}\ \bibnamefont {Zhang}},
  \bibinfo {author} {\bibfnamefont {Y.~J.}\ \bibnamefont {Zhou}}, \bibinfo
  {author} {\bibfnamefont {P.}~\bibnamefont {Chen}}, \bibinfo {author}
  {\bibfnamefont {C.~H.}\ \bibnamefont {Wan}}, \bibinfo {author} {\bibfnamefont
  {L.}~\bibnamefont {Han}}, \bibinfo {author} {\bibfnamefont {W.~X.}\
  \bibnamefont {Zhu}}, \bibinfo {author} {\bibfnamefont {S.~X.}\ \bibnamefont
  {Liang}}, \bibinfo {author} {\bibfnamefont {Y.~C.}\ \bibnamefont {Su}},
  \bibinfo {author} {\bibfnamefont {X.~F.}\ \bibnamefont {Han}}, \bibinfo
  {author} {\bibfnamefont {F.}~\bibnamefont {Pan}},\ and\ \bibinfo {author}
  {\bibfnamefont {C.}~\bibnamefont {Song}},\ }\bibfield  {title} {\bibinfo
  {title} {Efficient spin-to-charge conversion via altermagnetic spin splitting
  effect in antiferromagnet {R}u{O}$_2$},\ }\href
  {https://doi.org/10.1103/PhysRevLett.130.216701} {\bibfield  {journal}
  {\bibinfo  {journal} {Phys. Rev. Lett.}\ }\textbf {\bibinfo {volume} {130}},\
  \bibinfo {pages} {216701} (\bibinfo {year} {2023})}\BibitemShut {NoStop}%
\bibitem [{\citenamefont {Bai}\ \emph {et~al.}(2022)\citenamefont {Bai},
  \citenamefont {Han}, \citenamefont {Feng}, \citenamefont {Zhou},
  \citenamefont {Su}, \citenamefont {Wang}, \citenamefont {Liao}, \citenamefont
  {Zhu}, \citenamefont {Chen}, \citenamefont {Pan}, \citenamefont {Fan},\ and\
  \citenamefont {Song}}]{PhysRevLett.128.197202}%
  \BibitemOpen
  \bibfield  {author} {\bibinfo {author} {\bibfnamefont {H.}~\bibnamefont
  {Bai}}, \bibinfo {author} {\bibfnamefont {L.}~\bibnamefont {Han}}, \bibinfo
  {author} {\bibfnamefont {X.~Y.}\ \bibnamefont {Feng}}, \bibinfo {author}
  {\bibfnamefont {Y.~J.}\ \bibnamefont {Zhou}}, \bibinfo {author}
  {\bibfnamefont {R.~X.}\ \bibnamefont {Su}}, \bibinfo {author} {\bibfnamefont
  {Q.}~\bibnamefont {Wang}}, \bibinfo {author} {\bibfnamefont {L.~Y.}\
  \bibnamefont {Liao}}, \bibinfo {author} {\bibfnamefont {W.~X.}\ \bibnamefont
  {Zhu}}, \bibinfo {author} {\bibfnamefont {X.~Z.}\ \bibnamefont {Chen}},
  \bibinfo {author} {\bibfnamefont {F.}~\bibnamefont {Pan}}, \bibinfo {author}
  {\bibfnamefont {X.~L.}\ \bibnamefont {Fan}},\ and\ \bibinfo {author}
  {\bibfnamefont {C.}~\bibnamefont {Song}},\ }\bibfield  {title} {\bibinfo
  {title} {Observation of spin splitting torque in a collinear antiferromagnet
  {R}u{O}$_2$},\ }\href {https://doi.org/10.1103/PhysRevLett.128.197202}
  {\bibfield  {journal} {\bibinfo  {journal} {Phys. Rev. Lett.}\ }\textbf
  {\bibinfo {volume} {128}},\ \bibinfo {pages} {197202} (\bibinfo {year}
  {2022})}\BibitemShut {NoStop}%
\bibitem [{\citenamefont {Karube}\ \emph {et~al.}(2022)\citenamefont {Karube},
  \citenamefont {Tanaka}, \citenamefont {Sugawara}, \citenamefont {Kadoguchi},
  \citenamefont {Kohda},\ and\ \citenamefont {Nitta}}]{PhysRevLett.129.137201}%
  \BibitemOpen
  \bibfield  {author} {\bibinfo {author} {\bibfnamefont {S.}~\bibnamefont
  {Karube}}, \bibinfo {author} {\bibfnamefont {T.}~\bibnamefont {Tanaka}},
  \bibinfo {author} {\bibfnamefont {D.}~\bibnamefont {Sugawara}}, \bibinfo
  {author} {\bibfnamefont {N.}~\bibnamefont {Kadoguchi}}, \bibinfo {author}
  {\bibfnamefont {M.}~\bibnamefont {Kohda}},\ and\ \bibinfo {author}
  {\bibfnamefont {J.}~\bibnamefont {Nitta}},\ }\bibfield  {title} {\bibinfo
  {title} {Observation of spin-splitter torque in collinear antiferromagnetic
  {R}u{O}$_2$},\ }\href {https://doi.org/10.1103/PhysRevLett.129.137201}
  {\bibfield  {journal} {\bibinfo  {journal} {Phys. Rev. Lett.}\ }\textbf
  {\bibinfo {volume} {129}},\ \bibinfo {pages} {137201} (\bibinfo {year}
  {2022})}\BibitemShut {NoStop}%
\bibitem [{\citenamefont {Duan}\ \emph {et~al.}(2025)\citenamefont {Duan},
  \citenamefont {Zhang}, \citenamefont {Zhu}, \citenamefont {Liu},
  \citenamefont {Zhang}, \citenamefont {\ifmmode \check{Z}\else
  \v{Z}\fi{}uti\ifmmode~\acute{c}\else \'{c}\fi{}},\ and\ \citenamefont
  {Zhou}}]{PhysRevLett.134.106801}%
  \BibitemOpen
  \bibfield  {author} {\bibinfo {author} {\bibfnamefont {X.}~\bibnamefont
  {Duan}}, \bibinfo {author} {\bibfnamefont {J.}~\bibnamefont {Zhang}},
  \bibinfo {author} {\bibfnamefont {Z.}~\bibnamefont {Zhu}}, \bibinfo {author}
  {\bibfnamefont {Y.}~\bibnamefont {Liu}}, \bibinfo {author} {\bibfnamefont
  {Z.}~\bibnamefont {Zhang}}, \bibinfo {author} {\bibfnamefont
  {I.}~\bibnamefont {\ifmmode \check{Z}\else
  \v{Z}\fi{}uti\ifmmode~\acute{c}\else \'{c}\fi{}}},\ and\ \bibinfo {author}
  {\bibfnamefont {T.}~\bibnamefont {Zhou}},\ }\bibfield  {title} {\bibinfo
  {title} {Antiferroelectric altermagnets: Antiferroelectricity alters
  magnets},\ }\href {https://doi.org/10.1103/PhysRevLett.134.106801} {\bibfield
   {journal} {\bibinfo  {journal} {Phys. Rev. Lett.}\ }\textbf {\bibinfo
  {volume} {134}},\ \bibinfo {pages} {106801} (\bibinfo {year}
  {2025})}\BibitemShut {NoStop}%
\bibitem [{\citenamefont {Gu}\ \emph {et~al.}(2025)\citenamefont {Gu},
  \citenamefont {Liu}, \citenamefont {Zhu}, \citenamefont {Yananose},
  \citenamefont {Chen}, \citenamefont {Hu}, \citenamefont {Stroppa},\ and\
  \citenamefont {Liu}}]{PhysRevLett.134.106802}%
  \BibitemOpen
  \bibfield  {author} {\bibinfo {author} {\bibfnamefont {M.}~\bibnamefont
  {Gu}}, \bibinfo {author} {\bibfnamefont {Y.}~\bibnamefont {Liu}}, \bibinfo
  {author} {\bibfnamefont {H.}~\bibnamefont {Zhu}}, \bibinfo {author}
  {\bibfnamefont {K.}~\bibnamefont {Yananose}}, \bibinfo {author}
  {\bibfnamefont {X.}~\bibnamefont {Chen}}, \bibinfo {author} {\bibfnamefont
  {Y.}~\bibnamefont {Hu}}, \bibinfo {author} {\bibfnamefont {A.}~\bibnamefont
  {Stroppa}},\ and\ \bibinfo {author} {\bibfnamefont {Q.}~\bibnamefont {Liu}},\
  }\bibfield  {title} {\bibinfo {title} {Ferroelectric switchable
  altermagnetism},\ }\href {https://doi.org/10.1103/PhysRevLett.134.106802}
  {\bibfield  {journal} {\bibinfo  {journal} {Phys. Rev. Lett.}\ }\textbf
  {\bibinfo {volume} {134}},\ \bibinfo {pages} {106802} (\bibinfo {year}
  {2025})}\BibitemShut {NoStop}%
\bibitem [{\citenamefont {Werner}\ \emph {et~al.}(2024)\citenamefont {Werner},
  \citenamefont {Lysne},\ and\ \citenamefont {Murakami}}]{PhysRevB.110.235101}%
  \BibitemOpen
  \bibfield  {author} {\bibinfo {author} {\bibfnamefont {P.}~\bibnamefont
  {Werner}}, \bibinfo {author} {\bibfnamefont {M.}~\bibnamefont {Lysne}},\ and\
  \bibinfo {author} {\bibfnamefont {Y.}~\bibnamefont {Murakami}},\ }\bibfield
  {title} {\bibinfo {title} {High harmonic generation in altermagnets},\ }\href
  {https://doi.org/10.1103/PhysRevB.110.235101} {\bibfield  {journal} {\bibinfo
   {journal} {Phys. Rev. B}\ }\textbf {\bibinfo {volume} {110}},\ \bibinfo
  {pages} {235101} (\bibinfo {year} {2024})}\BibitemShut {NoStop}%
\bibitem [{\citenamefont {Ouassou}\ \emph {et~al.}(2023)\citenamefont
  {Ouassou}, \citenamefont {Brataas},\ and\ \citenamefont
  {Linder}}]{PhysRevLett.131.076003}%
  \BibitemOpen
  \bibfield  {author} {\bibinfo {author} {\bibfnamefont {J.~A.}\ \bibnamefont
  {Ouassou}}, \bibinfo {author} {\bibfnamefont {A.}~\bibnamefont {Brataas}},\
  and\ \bibinfo {author} {\bibfnamefont {J.}~\bibnamefont {Linder}},\
  }\bibfield  {title} {\bibinfo {title} {dc {J}osephson effect in
  altermagnets},\ }\href {https://doi.org/10.1103/PhysRevLett.131.076003}
  {\bibfield  {journal} {\bibinfo  {journal} {Phys. Rev. Lett.}\ }\textbf
  {\bibinfo {volume} {131}},\ \bibinfo {pages} {076003} (\bibinfo {year}
  {2023})}\BibitemShut {NoStop}%
\bibitem [{\citenamefont {Sun}\ \emph {et~al.}(2023)\citenamefont {Sun},
  \citenamefont {Brataas},\ and\ \citenamefont {Linder}}]{PhysRevB.108.054511}%
  \BibitemOpen
  \bibfield  {author} {\bibinfo {author} {\bibfnamefont {C.}~\bibnamefont
  {Sun}}, \bibinfo {author} {\bibfnamefont {A.}~\bibnamefont {Brataas}},\ and\
  \bibinfo {author} {\bibfnamefont {J.}~\bibnamefont {Linder}},\ }\bibfield
  {title} {\bibinfo {title} {Andreev reflection in altermagnets},\ }\href
  {https://doi.org/10.1103/PhysRevB.108.054511} {\bibfield  {journal} {\bibinfo
   {journal} {Phys. Rev. B}\ }\textbf {\bibinfo {volume} {108}},\ \bibinfo
  {pages} {054511} (\bibinfo {year} {2023})}\BibitemShut {NoStop}%
\bibitem [{\citenamefont {Papaj}(2023)}]{PhysRevB.108.L060508}%
  \BibitemOpen
  \bibfield  {author} {\bibinfo {author} {\bibfnamefont {M.}~\bibnamefont
  {Papaj}},\ }\bibfield  {title} {\bibinfo {title} {Andreev reflection at the
  altermagnet-superconductor interface},\ }\href
  {https://doi.org/10.1103/PhysRevB.108.L060508} {\bibfield  {journal}
  {\bibinfo  {journal} {Phys. Rev. B}\ }\textbf {\bibinfo {volume} {108}},\
  \bibinfo {pages} {L060508} (\bibinfo {year} {2023})}\BibitemShut {NoStop}%
\bibitem [{\citenamefont {Zhu}\ \emph {et~al.}(2023)\citenamefont {Zhu},
  \citenamefont {Zhuang}, \citenamefont {Wu},\ and\ \citenamefont
  {Yan}}]{PhysRevB.108.184505}%
  \BibitemOpen
  \bibfield  {author} {\bibinfo {author} {\bibfnamefont {D.}~\bibnamefont
  {Zhu}}, \bibinfo {author} {\bibfnamefont {Z.-Y.}\ \bibnamefont {Zhuang}},
  \bibinfo {author} {\bibfnamefont {Z.}~\bibnamefont {Wu}},\ and\ \bibinfo
  {author} {\bibfnamefont {Z.}~\bibnamefont {Yan}},\ }\bibfield  {title}
  {\bibinfo {title} {Topological superconductivity in two-dimensional
  altermagnetic metals},\ }\href {https://doi.org/10.1103/PhysRevB.108.184505}
  {\bibfield  {journal} {\bibinfo  {journal} {Phys. Rev. B}\ }\textbf {\bibinfo
  {volume} {108}},\ \bibinfo {pages} {184505} (\bibinfo {year}
  {2023})}\BibitemShut {NoStop}%
\bibitem [{\citenamefont {Brekke}\ \emph {et~al.}(2023)\citenamefont {Brekke},
  \citenamefont {Brataas},\ and\ \citenamefont
  {Sudb\o{}}}]{PhysRevB.108.224421}%
  \BibitemOpen
  \bibfield  {author} {\bibinfo {author} {\bibfnamefont {B.}~\bibnamefont
  {Brekke}}, \bibinfo {author} {\bibfnamefont {A.}~\bibnamefont {Brataas}},\
  and\ \bibinfo {author} {\bibfnamefont {A.}~\bibnamefont {Sudb\o{}}},\
  }\bibfield  {title} {\bibinfo {title} {Two-dimensional altermagnets:
  Superconductivity in a minimal microscopic model},\ }\href
  {https://doi.org/10.1103/PhysRevB.108.224421} {\bibfield  {journal} {\bibinfo
   {journal} {Phys. Rev. B}\ }\textbf {\bibinfo {volume} {108}},\ \bibinfo
  {pages} {224421} (\bibinfo {year} {2023})}\BibitemShut {NoStop}%
\bibitem [{\citenamefont {Zhang}\ \emph {et~al.}(2024)\citenamefont {Zhang},
  \citenamefont {Hu},\ and\ \citenamefont {Neupert}}]{Zhang2024}%
  \BibitemOpen
  \bibfield  {author} {\bibinfo {author} {\bibfnamefont {S.-B.}\ \bibnamefont
  {Zhang}}, \bibinfo {author} {\bibfnamefont {L.-H.}\ \bibnamefont {Hu}},\ and\
  \bibinfo {author} {\bibfnamefont {T.}~\bibnamefont {Neupert}},\ }\bibfield
  {title} {\bibinfo {title} {Finite-momentum {C}ooper pairing in proximitized
  altermagnets},\ }\href {https://doi.org/10.1038/s41467-024-45951-3}
  {\bibfield  {journal} {\bibinfo  {journal} {Nature Communications}\ }\textbf
  {\bibinfo {volume} {15}},\ \bibinfo {pages} {1801} (\bibinfo {year}
  {2024})}\BibitemShut {NoStop}%
\bibitem [{\citenamefont {Li}(2024)}]{PhysRevB.109.224502}%
  \BibitemOpen
  \bibfield  {author} {\bibinfo {author} {\bibfnamefont {Y.-X.}\ \bibnamefont
  {Li}},\ }\bibfield  {title} {\bibinfo {title} {Realizing tunable higher-order
  topological superconductors with altermagnets},\ }\href
  {https://doi.org/10.1103/PhysRevB.109.224502} {\bibfield  {journal} {\bibinfo
   {journal} {Phys. Rev. B}\ }\textbf {\bibinfo {volume} {109}},\ \bibinfo
  {pages} {224502} (\bibinfo {year} {2024})}\BibitemShut {NoStop}%
\bibitem [{\citenamefont {Li}\ and\ \citenamefont
  {Liu}(2023)}]{PhysRevB.108.205410}%
  \BibitemOpen
  \bibfield  {author} {\bibinfo {author} {\bibfnamefont {Y.-X.}\ \bibnamefont
  {Li}}\ and\ \bibinfo {author} {\bibfnamefont {C.-C.}\ \bibnamefont {Liu}},\
  }\bibfield  {title} {\bibinfo {title} {Majorana corner modes and tunable
  patterns in an altermagnet heterostructure},\ }\href
  {https://doi.org/10.1103/PhysRevB.108.205410} {\bibfield  {journal} {\bibinfo
   {journal} {Phys. Rev. B}\ }\textbf {\bibinfo {volume} {108}},\ \bibinfo
  {pages} {205410} (\bibinfo {year} {2023})}\BibitemShut {NoStop}%
\bibitem [{\citenamefont {Mazin}\ \emph {et~al.}(2021)\citenamefont {Mazin},
  \citenamefont {Koepernik}, \citenamefont {Johannes}, \citenamefont
  {González-Hernández},\ and\ \citenamefont
  {Šmejkal}}]{doi:10.1073/pnas.2108924118}%
  \BibitemOpen
  \bibfield  {author} {\bibinfo {author} {\bibfnamefont {I.~I.}\ \bibnamefont
  {Mazin}}, \bibinfo {author} {\bibfnamefont {K.}~\bibnamefont {Koepernik}},
  \bibinfo {author} {\bibfnamefont {M.~D.}\ \bibnamefont {Johannes}}, \bibinfo
  {author} {\bibfnamefont {R.}~\bibnamefont {González-Hernández}},\ and\
  \bibinfo {author} {\bibfnamefont {L.}~\bibnamefont {Šmejkal}},\ }\bibfield
  {title} {\bibinfo {title} {Prediction of unconventional magnetism in doped
  {F}e{S}b$_2$},\ }\href {https://doi.org/10.1073/pnas.2108924118} {\bibfield
  {journal} {\bibinfo  {journal} {Proceedings of the National Academy of
  Sciences}\ }\textbf {\bibinfo {volume} {118}},\ \bibinfo {pages}
  {e2108924118} (\bibinfo {year} {2021})}\BibitemShut {NoStop}%
\bibitem [{\citenamefont {Feng}\ \emph {et~al.}(2024)\citenamefont {Feng},
  \citenamefont {Bai}, \citenamefont {Fan}, \citenamefont {Guo}, \citenamefont
  {Zhang}, \citenamefont {Chai}, \citenamefont {Wang}, \citenamefont {Xue},
  \citenamefont {Song},\ and\ \citenamefont {Fan}}]{PhysRevLett.132.086701}%
  \BibitemOpen
  \bibfield  {author} {\bibinfo {author} {\bibfnamefont {X.}~\bibnamefont
  {Feng}}, \bibinfo {author} {\bibfnamefont {H.}~\bibnamefont {Bai}}, \bibinfo
  {author} {\bibfnamefont {X.}~\bibnamefont {Fan}}, \bibinfo {author}
  {\bibfnamefont {M.}~\bibnamefont {Guo}}, \bibinfo {author} {\bibfnamefont
  {Z.}~\bibnamefont {Zhang}}, \bibinfo {author} {\bibfnamefont
  {G.}~\bibnamefont {Chai}}, \bibinfo {author} {\bibfnamefont {T.}~\bibnamefont
  {Wang}}, \bibinfo {author} {\bibfnamefont {D.}~\bibnamefont {Xue}}, \bibinfo
  {author} {\bibfnamefont {C.}~\bibnamefont {Song}},\ and\ \bibinfo {author}
  {\bibfnamefont {X.}~\bibnamefont {Fan}},\ }\bibfield  {title} {\bibinfo
  {title} {Incommensurate spin density wave in antiferromagnetic {R}u{O}$_2$
  evinced by abnormal spin splitting torque},\ }\href
  {https://doi.org/10.1103/PhysRevLett.132.086701} {\bibfield  {journal}
  {\bibinfo  {journal} {Phys. Rev. Lett.}\ }\textbf {\bibinfo {volume} {132}},\
  \bibinfo {pages} {086701} (\bibinfo {year} {2024})}\BibitemShut {NoStop}%
\bibitem [{\citenamefont {McClarty}\ and\ \citenamefont
  {Rau}(2024)}]{PhysRevLett.132.176702}%
  \BibitemOpen
  \bibfield  {author} {\bibinfo {author} {\bibfnamefont {P.~A.}\ \bibnamefont
  {McClarty}}\ and\ \bibinfo {author} {\bibfnamefont {J.~G.}\ \bibnamefont
  {Rau}},\ }\bibfield  {title} {\bibinfo {title} {Landau theory of
  altermagnetism},\ }\href {https://doi.org/10.1103/PhysRevLett.132.176702}
  {\bibfield  {journal} {\bibinfo  {journal} {Phys. Rev. Lett.}\ }\textbf
  {\bibinfo {volume} {132}},\ \bibinfo {pages} {176702} (\bibinfo {year}
  {2024})}\BibitemShut {NoStop}%
\bibitem [{\citenamefont {Feng}\ \emph {et~al.}(2022)\citenamefont {Feng},
  \citenamefont {Zhou}, \citenamefont {{\v{S}}mejkal}, \citenamefont {Wu},
  \citenamefont {Zhu}, \citenamefont {Guo}, \citenamefont
  {Gonz{\'a}lez-Hern{\'a}ndez}, \citenamefont {Wang}, \citenamefont {Yan},
  \citenamefont {Qin}, \citenamefont {Zhang}, \citenamefont {Wu}, \citenamefont
  {Chen}, \citenamefont {Meng}, \citenamefont {Liu}, \citenamefont {Xia},
  \citenamefont {Sinova}, \citenamefont {Jungwirth},\ and\ \citenamefont
  {Liu}}]{Feng2022}%
  \BibitemOpen
  \bibfield  {author} {\bibinfo {author} {\bibfnamefont {Z.}~\bibnamefont
  {Feng}}, \bibinfo {author} {\bibfnamefont {X.}~\bibnamefont {Zhou}}, \bibinfo
  {author} {\bibfnamefont {L.}~\bibnamefont {{\v{S}}mejkal}}, \bibinfo {author}
  {\bibfnamefont {L.}~\bibnamefont {Wu}}, \bibinfo {author} {\bibfnamefont
  {Z.}~\bibnamefont {Zhu}}, \bibinfo {author} {\bibfnamefont {H.}~\bibnamefont
  {Guo}}, \bibinfo {author} {\bibfnamefont {R.}~\bibnamefont
  {Gonz{\'a}lez-Hern{\'a}ndez}}, \bibinfo {author} {\bibfnamefont
  {X.}~\bibnamefont {Wang}}, \bibinfo {author} {\bibfnamefont {H.}~\bibnamefont
  {Yan}}, \bibinfo {author} {\bibfnamefont {P.}~\bibnamefont {Qin}}, \bibinfo
  {author} {\bibfnamefont {X.}~\bibnamefont {Zhang}}, \bibinfo {author}
  {\bibfnamefont {H.}~\bibnamefont {Wu}}, \bibinfo {author} {\bibfnamefont
  {H.}~\bibnamefont {Chen}}, \bibinfo {author} {\bibfnamefont {Z.}~\bibnamefont
  {Meng}}, \bibinfo {author} {\bibfnamefont {L.}~\bibnamefont {Liu}}, \bibinfo
  {author} {\bibfnamefont {Z.}~\bibnamefont {Xia}}, \bibinfo {author}
  {\bibfnamefont {J.}~\bibnamefont {Sinova}}, \bibinfo {author} {\bibfnamefont
  {T.}~\bibnamefont {Jungwirth}},\ and\ \bibinfo {author} {\bibfnamefont
  {Z.}~\bibnamefont {Liu}},\ }\bibfield  {title} {\bibinfo {title} {An
  anomalous {H}all effect in altermagnetic ruthenium dioxide},\ }\href
  {https://doi.org/10.1038/s41928-022-00866-z} {\bibfield  {journal} {\bibinfo
  {journal} {Nature Electronics}\ }\textbf {\bibinfo {volume} {5}},\ \bibinfo
  {pages} {735} (\bibinfo {year} {2022})}\BibitemShut {NoStop}%
\bibitem [{\citenamefont {Šmejkal}\ \emph {et~al.}(2020)\citenamefont
  {Šmejkal}, \citenamefont {González-Hernández}, \citenamefont {Jungwirth},\
  and\ \citenamefont {Sinova}}]{doi:10.1126/sciadv.aaz8809}%
  \BibitemOpen
  \bibfield  {author} {\bibinfo {author} {\bibfnamefont {L.}~\bibnamefont
  {Šmejkal}}, \bibinfo {author} {\bibfnamefont {R.}~\bibnamefont
  {González-Hernández}}, \bibinfo {author} {\bibfnamefont {T.}~\bibnamefont
  {Jungwirth}},\ and\ \bibinfo {author} {\bibfnamefont {J.}~\bibnamefont
  {Sinova}},\ }\bibfield  {title} {\bibinfo {title} {Crystal time-reversal
  symmetry breaking and spontaneous {H}all effect in collinear
  antiferromagnets},\ }\href {https://doi.org/10.1126/sciadv.aaz8809}
  {\bibfield  {journal} {\bibinfo  {journal} {Science Advances}\ }\textbf
  {\bibinfo {volume} {6}},\ \bibinfo {pages} {eaaz8809} (\bibinfo {year}
  {2020})}\BibitemShut {NoStop}%
\bibitem [{\citenamefont {Weber}\ \emph {et~al.}(2024)\citenamefont {Weber},
  \citenamefont {Wust}, \citenamefont {Haag}, \citenamefont {Akashdeep},
  \citenamefont {Leckron}, \citenamefont {Schmitt}, \citenamefont {Ramos},
  \citenamefont {Kikkawa}, \citenamefont {Saitoh}, \citenamefont {Kläui},
  \citenamefont {Šmejkal}, \citenamefont {Sinova}, \citenamefont
  {Aeschlimann}, \citenamefont {Jakob}, \citenamefont {Stadtmüller},\ and\
  \citenamefont {Schneider}}]{weber2024opticalexcitationspinpolarization}%
  \BibitemOpen
  \bibfield  {author} {\bibinfo {author} {\bibfnamefont {M.}~\bibnamefont
  {Weber}}, \bibinfo {author} {\bibfnamefont {S.}~\bibnamefont {Wust}},
  \bibinfo {author} {\bibfnamefont {L.}~\bibnamefont {Haag}}, \bibinfo {author}
  {\bibfnamefont {A.}~\bibnamefont {Akashdeep}}, \bibinfo {author}
  {\bibfnamefont {K.}~\bibnamefont {Leckron}}, \bibinfo {author} {\bibfnamefont
  {C.}~\bibnamefont {Schmitt}}, \bibinfo {author} {\bibfnamefont
  {R.}~\bibnamefont {Ramos}}, \bibinfo {author} {\bibfnamefont
  {T.}~\bibnamefont {Kikkawa}}, \bibinfo {author} {\bibfnamefont
  {E.}~\bibnamefont {Saitoh}}, \bibinfo {author} {\bibfnamefont
  {M.}~\bibnamefont {Kläui}}, \bibinfo {author} {\bibfnamefont
  {L.}~\bibnamefont {Šmejkal}}, \bibinfo {author} {\bibfnamefont
  {J.}~\bibnamefont {Sinova}}, \bibinfo {author} {\bibfnamefont
  {M.}~\bibnamefont {Aeschlimann}}, \bibinfo {author} {\bibfnamefont
  {G.}~\bibnamefont {Jakob}}, \bibinfo {author} {\bibfnamefont
  {B.}~\bibnamefont {Stadtmüller}},\ and\ \bibinfo {author} {\bibfnamefont
  {H.~C.}\ \bibnamefont {Schneider}},\ }\bibfield  {title} {\bibinfo {title}
  {All optical excitation of spin polarization in $d$-wave altermagnets},\
  }\href {https://arxiv.org/abs/2408.05187} {\  (\bibinfo {year} {2024})},\
  \Eprint {https://arxiv.org/abs/2408.05187} {arXiv:2408.05187} \BibitemShut
  {NoStop}%
\bibitem [{\citenamefont {Lee}\ \emph {et~al.}(2024)\citenamefont {Lee},
  \citenamefont {Lee}, \citenamefont {Jung}, \citenamefont {Jung},
  \citenamefont {Kim}, \citenamefont {Lee}, \citenamefont {Seok}, \citenamefont
  {Kim}, \citenamefont {Park}, \citenamefont {\ifmmode~\check{S}\else
  \v{S}\fi{}mejkal}, \citenamefont {Kang},\ and\ \citenamefont
  {Kim}}]{PhysRevLett.132.036702}%
  \BibitemOpen
  \bibfield  {author} {\bibinfo {author} {\bibfnamefont {S.}~\bibnamefont
  {Lee}}, \bibinfo {author} {\bibfnamefont {S.}~\bibnamefont {Lee}}, \bibinfo
  {author} {\bibfnamefont {S.}~\bibnamefont {Jung}}, \bibinfo {author}
  {\bibfnamefont {J.}~\bibnamefont {Jung}}, \bibinfo {author} {\bibfnamefont
  {D.}~\bibnamefont {Kim}}, \bibinfo {author} {\bibfnamefont {Y.}~\bibnamefont
  {Lee}}, \bibinfo {author} {\bibfnamefont {B.}~\bibnamefont {Seok}}, \bibinfo
  {author} {\bibfnamefont {J.}~\bibnamefont {Kim}}, \bibinfo {author}
  {\bibfnamefont {B.~G.}\ \bibnamefont {Park}}, \bibinfo {author}
  {\bibfnamefont {L.}~\bibnamefont {\ifmmode~\check{S}\else \v{S}\fi{}mejkal}},
  \bibinfo {author} {\bibfnamefont {C.-J.}\ \bibnamefont {Kang}},\ and\
  \bibinfo {author} {\bibfnamefont {C.}~\bibnamefont {Kim}},\ }\bibfield
  {title} {\bibinfo {title} {Broken {K}ramers degeneracy in altermagnetic
  {M}n{T}e},\ }\href {https://doi.org/10.1103/PhysRevLett.132.036702}
  {\bibfield  {journal} {\bibinfo  {journal} {Phys. Rev. Lett.}\ }\textbf
  {\bibinfo {volume} {132}},\ \bibinfo {pages} {036702} (\bibinfo {year}
  {2024})}\BibitemShut {NoStop}%
\bibitem [{\citenamefont {Mazin}(2023)}]{PhysRevB.107.L100418}%
  \BibitemOpen
  \bibfield  {author} {\bibinfo {author} {\bibfnamefont {I.~I.}\ \bibnamefont
  {Mazin}},\ }\bibfield  {title} {\bibinfo {title} {Altermagnetism in {M}n{T}e:
  Origin, predicted manifestations, and routes to detwinning},\ }\href
  {https://doi.org/10.1103/PhysRevB.107.L100418} {\bibfield  {journal}
  {\bibinfo  {journal} {Phys. Rev. B}\ }\textbf {\bibinfo {volume} {107}},\
  \bibinfo {pages} {L100418} (\bibinfo {year} {2023})}\BibitemShut {NoStop}%
\bibitem [{\citenamefont {Faria~Junior}\ \emph {et~al.}(2023)\citenamefont
  {Faria~Junior}, \citenamefont {de~Mare}, \citenamefont {Zollner},
  \citenamefont {Ahn}, \citenamefont {Erlingsson}, \citenamefont {van
  Schilfgaarde},\ and\ \citenamefont {V\'yborn\'y}}]{PhysRevB.107.L100417}%
  \BibitemOpen
  \bibfield  {author} {\bibinfo {author} {\bibfnamefont {P.~E.}\ \bibnamefont
  {Faria~Junior}}, \bibinfo {author} {\bibfnamefont {K.~A.}\ \bibnamefont
  {de~Mare}}, \bibinfo {author} {\bibfnamefont {K.}~\bibnamefont {Zollner}},
  \bibinfo {author} {\bibfnamefont {K.-h.}\ \bibnamefont {Ahn}}, \bibinfo
  {author} {\bibfnamefont {S.~I.}\ \bibnamefont {Erlingsson}}, \bibinfo
  {author} {\bibfnamefont {M.}~\bibnamefont {van Schilfgaarde}},\ and\ \bibinfo
  {author} {\bibfnamefont {K.}~\bibnamefont {V\'yborn\'y}},\ }\bibfield
  {title} {\bibinfo {title} {Sensitivity of the {M}n{T}e valence band to the
  orientation of magnetic moments},\ }\href
  {https://doi.org/10.1103/PhysRevB.107.L100417} {\bibfield  {journal}
  {\bibinfo  {journal} {Phys. Rev. B}\ }\textbf {\bibinfo {volume} {107}},\
  \bibinfo {pages} {L100417} (\bibinfo {year} {2023})}\BibitemShut {NoStop}%
\bibitem [{\citenamefont {Reichlová}\ \emph {et~al.}(2021)\citenamefont
  {Reichlová}, \citenamefont {Seeger}, \citenamefont {González-Hernández},
  \citenamefont {Kounta}, \citenamefont {Schlitz}, \citenamefont {Kriegner},
  \citenamefont {Ritzinger}, \citenamefont {Lammel}, \citenamefont {Leiviskä},
  \citenamefont {Petříček}, \citenamefont {Doležal}, \citenamefont
  {Schmoranzerová}, \citenamefont {Bad'ura}, \citenamefont {Thomas},
  \citenamefont {Baltz}, \citenamefont {Michez}, \citenamefont {Sinova},
  \citenamefont {Goennenwein}, \citenamefont {Jungwirth},\ and\ \citenamefont
  {Šmejkal}}]{reichlova2021macroscopictimereversalsymmetry}%
  \BibitemOpen
  \bibfield  {author} {\bibinfo {author} {\bibfnamefont {H.}~\bibnamefont
  {Reichlová}}, \bibinfo {author} {\bibfnamefont {R.~L.}\ \bibnamefont
  {Seeger}}, \bibinfo {author} {\bibfnamefont {R.}~\bibnamefont
  {González-Hernández}}, \bibinfo {author} {\bibfnamefont {I.}~\bibnamefont
  {Kounta}}, \bibinfo {author} {\bibfnamefont {R.}~\bibnamefont {Schlitz}},
  \bibinfo {author} {\bibfnamefont {D.}~\bibnamefont {Kriegner}}, \bibinfo
  {author} {\bibfnamefont {P.}~\bibnamefont {Ritzinger}}, \bibinfo {author}
  {\bibfnamefont {M.}~\bibnamefont {Lammel}}, \bibinfo {author} {\bibfnamefont
  {M.}~\bibnamefont {Leiviskä}}, \bibinfo {author} {\bibfnamefont
  {V.}~\bibnamefont {Petříček}}, \bibinfo {author} {\bibfnamefont
  {P.}~\bibnamefont {Doležal}}, \bibinfo {author} {\bibfnamefont
  {E.}~\bibnamefont {Schmoranzerová}}, \bibinfo {author} {\bibfnamefont
  {A.}~\bibnamefont {Bad'ura}}, \bibinfo {author} {\bibfnamefont
  {A.}~\bibnamefont {Thomas}}, \bibinfo {author} {\bibfnamefont
  {V.}~\bibnamefont {Baltz}}, \bibinfo {author} {\bibfnamefont
  {L.}~\bibnamefont {Michez}}, \bibinfo {author} {\bibfnamefont
  {J.}~\bibnamefont {Sinova}}, \bibinfo {author} {\bibfnamefont {S.~T.~B.}\
  \bibnamefont {Goennenwein}}, \bibinfo {author} {\bibfnamefont
  {T.}~\bibnamefont {Jungwirth}},\ and\ \bibinfo {author} {\bibfnamefont
  {L.}~\bibnamefont {Šmejkal}},\ }\bibfield  {title} {\bibinfo {title}
  {Macroscopic time reversal symmetry breaking by staggered spin-momentum
  interaction},\ }\href {https://arxiv.org/abs/2012.15651} {\  (\bibinfo {year}
  {2021})},\ \Eprint {https://arxiv.org/abs/2012.15651} {arXiv:2012.15651}
  \BibitemShut {NoStop}%
\bibitem [{\citenamefont {Kasuya}(1956)}]{10.1143/PTP.16.45}%
  \BibitemOpen
  \bibfield  {author} {\bibinfo {author} {\bibfnamefont {T.}~\bibnamefont
  {Kasuya}},\ }\bibfield  {title} {\bibinfo {title} {{A Theory of Metallic
  Ferro- and Antiferromagnetism on Zener's Model}},\ }\href
  {https://doi.org/10.1143/PTP.16.45} {\bibfield  {journal} {\bibinfo
  {journal} {Progress of Theoretical Physics}\ }\textbf {\bibinfo {volume}
  {16}},\ \bibinfo {pages} {45} (\bibinfo {year} {1956})}\BibitemShut {NoStop}%
\bibitem [{\citenamefont {Yosida}(1957)}]{PhysRev.106.893}%
  \BibitemOpen
  \bibfield  {author} {\bibinfo {author} {\bibfnamefont {K.}~\bibnamefont
  {Yosida}},\ }\bibfield  {title} {\bibinfo {title} {Magnetic properties of
  {C}u-{M}n alloys},\ }\href {https://doi.org/10.1103/PhysRev.106.893}
  {\bibfield  {journal} {\bibinfo  {journal} {Phys. Rev.}\ }\textbf {\bibinfo
  {volume} {106}},\ \bibinfo {pages} {893} (\bibinfo {year}
  {1957})}\BibitemShut {NoStop}%
\bibitem [{\citenamefont {Ruderman}\ and\ \citenamefont
  {Kittel}(1954)}]{PhysRev.96.99}%
  \BibitemOpen
  \bibfield  {author} {\bibinfo {author} {\bibfnamefont {M.~A.}\ \bibnamefont
  {Ruderman}}\ and\ \bibinfo {author} {\bibfnamefont {C.}~\bibnamefont
  {Kittel}},\ }\bibfield  {title} {\bibinfo {title} {Indirect exchange coupling
  of nuclear magnetic moments by conduction electrons},\ }\href
  {https://doi.org/10.1103/PhysRev.96.99} {\bibfield  {journal} {\bibinfo
  {journal} {Phys. Rev.}\ }\textbf {\bibinfo {volume} {96}},\ \bibinfo {pages}
  {99} (\bibinfo {year} {1954})}\BibitemShut {NoStop}%
\bibitem [{\citenamefont
  {Lee}(2024)}]{lee2024magneticimpuritiesaltermagneticmetal}%
  \BibitemOpen
  \bibfield  {author} {\bibinfo {author} {\bibfnamefont {Y.-L.}\ \bibnamefont
  {Lee}},\ }\bibfield  {title} {\bibinfo {title} {Magnetic impurities in an
  altermagnetic metal},\ }\href {https://arxiv.org/abs/2312.15733} {\
  (\bibinfo {year} {2024})},\ \Eprint {https://arxiv.org/abs/2312.15733}
  {arXiv:2312.15733} \BibitemShut {NoStop}%
\bibitem [{\citenamefont {Amundsen}\ \emph {et~al.}(2024)\citenamefont
  {Amundsen}, \citenamefont {Brataas},\ and\ \citenamefont
  {Linder}}]{PhysRevB.110.054427}%
  \BibitemOpen
  \bibfield  {author} {\bibinfo {author} {\bibfnamefont {M.}~\bibnamefont
  {Amundsen}}, \bibinfo {author} {\bibfnamefont {A.}~\bibnamefont {Brataas}},\
  and\ \bibinfo {author} {\bibfnamefont {J.}~\bibnamefont {Linder}},\
  }\bibfield  {title} {\bibinfo {title} {{RKKY} interaction in {R}ashba
  altermagnets},\ }\href {https://doi.org/10.1103/PhysRevB.110.054427}
  {\bibfield  {journal} {\bibinfo  {journal} {Phys. Rev. B}\ }\textbf {\bibinfo
  {volume} {110}},\ \bibinfo {pages} {054427} (\bibinfo {year}
  {2024})}\BibitemShut {NoStop}%
\bibitem [{\citenamefont {Yarmohammadi}\ \emph {et~al.}(2025)\citenamefont
  {Yarmohammadi}, \citenamefont {Z\"ulicke}, \citenamefont {Berakdar},
  \citenamefont {Linder},\ and\ \citenamefont {Freericks}}]{k3xb-8pts}%
  \BibitemOpen
  \bibfield  {author} {\bibinfo {author} {\bibfnamefont {M.}~\bibnamefont
  {Yarmohammadi}}, \bibinfo {author} {\bibfnamefont {U.}~\bibnamefont
  {Z\"ulicke}}, \bibinfo {author} {\bibfnamefont {J.}~\bibnamefont {Berakdar}},
  \bibinfo {author} {\bibfnamefont {J.}~\bibnamefont {Linder}},\ and\ \bibinfo
  {author} {\bibfnamefont {J.~K.}\ \bibnamefont {Freericks}},\ }\bibfield
  {title} {\bibinfo {title} {Anisotropic light-tailored {RKKY} interaction in
  two-dimensional $d$-wave altermagnets},\ }\href
  {https://doi.org/10.1103/k3xb-8pts} {\bibfield  {journal} {\bibinfo
  {journal} {Phys. Rev. B}\ }\textbf {\bibinfo {volume} {111}},\ \bibinfo
  {pages} {224412} (\bibinfo {year} {2025})}\BibitemShut {NoStop}%
\bibitem [{\citenamefont
  {Kundu}(2025)}]{kundu2025rkkyinteractionmediatedspinpolarized}%
  \BibitemOpen
  \bibfield  {author} {\bibinfo {author} {\bibfnamefont {A.}~\bibnamefont
  {Kundu}},\ }\bibfield  {title} {\bibinfo {title} {{RKKY} interaction mediated
  by a spin-polarized 2d electron gas with {R}ashba and altermagnetic
  coupling},\ }\href {https://arxiv.org/abs/2509.10778} {\  (\bibinfo {year}
  {2025})},\ \Eprint {https://arxiv.org/abs/2509.10778} {arXiv:2509.10778}
  \BibitemShut {NoStop}%
\bibitem [{\citenamefont {Yarmohammadi}\ \emph
  {et~al.}(2026{\natexlab{b}})\citenamefont {Yarmohammadi}, \citenamefont
  {Fu},\ and\ \citenamefont
  {Freericks}}]{yarmohammadi2026efficienttwocolorfloquetcontrol}%
  \BibitemOpen
  \bibfield  {author} {\bibinfo {author} {\bibfnamefont {M.}~\bibnamefont
  {Yarmohammadi}}, \bibinfo {author} {\bibfnamefont {P.-H.}\ \bibnamefont
  {Fu}},\ and\ \bibinfo {author} {\bibfnamefont {J.~K.}\ \bibnamefont
  {Freericks}},\ }\bibfield  {title} {\bibinfo {title} {Efficient two-color
  {F}loquet control of the {RKKY} interaction in altermagnets},\ }\href
  {https://arxiv.org/abs/2602.20862} {\  (\bibinfo {year}
  {2026}{\natexlab{b}})},\ \Eprint {https://arxiv.org/abs/2602.20862}
  {arXiv:2602.20862 [cond-mat.mes-hall]} \BibitemShut {NoStop}%
\bibitem [{\citenamefont {Duan}\ \emph {et~al.}(2026)\citenamefont {Duan},
  \citenamefont {Fang}, \citenamefont {Deng}, \citenamefont {Wang},\ and\
  \citenamefont {Yang}}]{gr5n-174c}%
  \BibitemOpen
  \bibfield  {author} {\bibinfo {author} {\bibfnamefont {H.-J.}\ \bibnamefont
  {Duan}}, \bibinfo {author} {\bibfnamefont {M.-S.}\ \bibnamefont {Fang}},
  \bibinfo {author} {\bibfnamefont {M.-X.}\ \bibnamefont {Deng}}, \bibinfo
  {author} {\bibfnamefont {R.}~\bibnamefont {Wang}},\ and\ \bibinfo {author}
  {\bibfnamefont {M.}~\bibnamefont {Yang}},\ }\bibfield  {title} {\bibinfo
  {title} {N\'eel vector and {R}ashba spin-orbit coupling effects on {RKKY}
  interaction in two-dimensional $d$-wave altermagnets},\ }\href
  {https://doi.org/10.1103/gr5n-174c} {\bibfield  {journal} {\bibinfo
  {journal} {Phys. Rev. B}\ }\textbf {\bibinfo {volume} {113}},\ \bibinfo
  {pages} {075104} (\bibinfo {year} {2026})}\BibitemShut {NoStop}%
\bibitem [{\citenamefont {Giustino}(2017)}]{giustino2016electron}%
  \BibitemOpen
  \bibfield  {author} {\bibinfo {author} {\bibfnamefont {F.}~\bibnamefont
  {Giustino}},\ }\bibfield  {title} {\bibinfo {title} {Electron-phonon
  interactions from first principles},\ }\href
  {https://doi.org/10.1103/RevModPhys.89.015003} {\bibfield  {journal}
  {\bibinfo  {journal} {Rev. Mod. Phys.}\ }\textbf {\bibinfo {volume} {89}},\
  \bibinfo {pages} {015003} (\bibinfo {year} {2017})}\BibitemShut {NoStop}%
\bibitem [{\citenamefont {Iorsh}(2025)}]{6wxh-p4mc}%
  \BibitemOpen
  \bibfield  {author} {\bibinfo {author} {\bibfnamefont {I.~V.}\ \bibnamefont
  {Iorsh}},\ }\bibfield  {title} {\bibinfo {title} {Electron pairing by
  dispersive phonons in altermagnets: Reentrant superconductivity and
  continuous transition to finite momentum superconducting state},\ }\href
  {https://doi.org/10.1103/6wxh-p4mc} {\bibfield  {journal} {\bibinfo
  {journal} {Phys. Rev. B}\ }\textbf {\bibinfo {volume} {111}},\ \bibinfo
  {pages} {L220503} (\bibinfo {year} {2025})}\BibitemShut {NoStop}%
\bibitem [{\citenamefont {Leraand}\ \emph {et~al.}(2025)\citenamefont
  {Leraand}, \citenamefont {M\ae{}land},\ and\ \citenamefont
  {Sudb\o{}}}]{leraand2025phononmediatedspinpolarizedsuperconductivityaltermagnets}%
  \BibitemOpen
  \bibfield  {author} {\bibinfo {author} {\bibfnamefont {K.}~\bibnamefont
  {Leraand}}, \bibinfo {author} {\bibfnamefont {K.}~\bibnamefont
  {M\ae{}land}},\ and\ \bibinfo {author} {\bibfnamefont {A.}~\bibnamefont
  {Sudb\o{}}},\ }\bibfield  {title} {\bibinfo {title} {Phonon-mediated
  spin-polarized superconductivity in altermagnets},\ }\href
  {https://doi.org/10.1103/g4dl-1ff2} {\bibfield  {journal} {\bibinfo
  {journal} {Phys. Rev. B}\ }\textbf {\bibinfo {volume} {112}},\ \bibinfo
  {pages} {104510} (\bibinfo {year} {2025})}\BibitemShut {NoStop}%
\bibitem [{\citenamefont {Steward}\ \emph {et~al.}(2023)\citenamefont
  {Steward}, \citenamefont {Fernandes},\ and\ \citenamefont
  {Schmalian}}]{nlpj-1dt5}%
  \BibitemOpen
  \bibfield  {author} {\bibinfo {author} {\bibfnamefont {C.~R.~W.}\
  \bibnamefont {Steward}}, \bibinfo {author} {\bibfnamefont {R.~M.}\
  \bibnamefont {Fernandes}},\ and\ \bibinfo {author} {\bibfnamefont
  {J.}~\bibnamefont {Schmalian}},\ }\bibfield  {title} {\bibinfo {title}
  {Dynamic paramagnon-polarons in altermagnets},\ }\href
  {https://doi.org/10.1103/PhysRevB.108.144418} {\bibfield  {journal} {\bibinfo
   {journal} {Phys. Rev. B}\ }\textbf {\bibinfo {volume} {108}},\ \bibinfo
  {pages} {144418} (\bibinfo {year} {2023})}\BibitemShut {NoStop}%
\bibitem [{\citenamefont {Hodt}\ \emph {et~al.}(2026)\citenamefont {Hodt},
  \citenamefont {Qaiumzadeh},\ and\ \citenamefont
  {Linder}}]{hodt2025phononenhancedopticalspinconductivityspinsplitter}%
  \BibitemOpen
  \bibfield  {author} {\bibinfo {author} {\bibfnamefont {E.~W.}\ \bibnamefont
  {Hodt}}, \bibinfo {author} {\bibfnamefont {A.}~\bibnamefont {Qaiumzadeh}},\
  and\ \bibinfo {author} {\bibfnamefont {J.}~\bibnamefont {Linder}},\
  }\bibfield  {title} {\bibinfo {title} {Phonon-enhanced optical spin
  conductivity and spin-splitter effect in altermagnets},\ }\href
  {https://doi.org/10.1103/zwz9-l7wf} {\bibfield  {journal} {\bibinfo
  {journal} {Phys. Rev. B}\ }\textbf {\bibinfo {volume} {113}},\ \bibinfo
  {pages} {054403} (\bibinfo {year} {2026})}\BibitemShut {NoStop}%
\bibitem [{\citenamefont {Yarmohammadi}\ \emph
  {et~al.}(2026{\natexlab{c}})\citenamefont {Yarmohammadi}, \citenamefont
  {Linder},\ and\ \citenamefont
  {Freericks}}]{yarmohammadi2025slowphononcontrolspinedelstein}%
  \BibitemOpen
  \bibfield  {author} {\bibinfo {author} {\bibfnamefont {M.}~\bibnamefont
  {Yarmohammadi}}, \bibinfo {author} {\bibfnamefont {J.}~\bibnamefont
  {Linder}},\ and\ \bibinfo {author} {\bibfnamefont {J.~K.}\ \bibnamefont
  {Freericks}},\ }\bibfield  {title} {\bibinfo {title} {Slow-phonon control of
  spin {E}delstein effect in {R}ashba $d$-wave altermagnets},\ }\href
  {https://doi.org/10.1103/4txy-gy8f} {\bibfield  {journal} {\bibinfo
  {journal} {Phys. Rev. B}\ ,\ } (\bibinfo {year}
  {2026}{\natexlab{c}})}\BibitemShut {NoStop}%
\bibitem [{\citenamefont {Winkler}(2003)}]{winkler2003spin}%
  \BibitemOpen
  \bibfield  {author} {\bibinfo {author} {\bibfnamefont {R.}~\bibnamefont
  {Winkler}},\ }\href {https://books.google.se/books?id=LQhcSCuzC3IC} {\emph
  {\bibinfo {title} {Spin-orbit Coupling Effects in Two-Dimensional Electron
  and Hole Systems}}},\ Springer Tracts in Modern Physics\ (\bibinfo
  {publisher} {Springer Berlin Heidelberg},\ \bibinfo {year}
  {2003})\BibitemShut {NoStop}%
\bibitem [{\citenamefont {Yarmohammadi}\ \emph
  {et~al.}(2026{\natexlab{d}})\citenamefont {Yarmohammadi}, \citenamefont
  {Berritta}, \citenamefont {Bukov}, \citenamefont {\ifmmode~\check{S}\else
  \v{S}\fi{}mejkal}, \citenamefont {Linder},\ and\ \citenamefont
  {Oppeneer}}]{yarmohammadi2025spinpolarizationengineeringdwave}%
  \BibitemOpen
  \bibfield  {author} {\bibinfo {author} {\bibfnamefont {M.}~\bibnamefont
  {Yarmohammadi}}, \bibinfo {author} {\bibfnamefont {M.}~\bibnamefont
  {Berritta}}, \bibinfo {author} {\bibfnamefont {M.}~\bibnamefont {Bukov}},
  \bibinfo {author} {\bibfnamefont {L.}~\bibnamefont {\ifmmode~\check{S}\else
  \v{S}\fi{}mejkal}}, \bibinfo {author} {\bibfnamefont {J.}~\bibnamefont
  {Linder}},\ and\ \bibinfo {author} {\bibfnamefont {P.~M.}\ \bibnamefont
  {Oppeneer}},\ }\bibfield  {title} {\bibinfo {title} {Spin polarization
  engineering in $d$-wave altermagnets},\ }\href
  {https://doi.org/10.1103/xt23-9pnv} {\bibfield  {journal} {\bibinfo
  {journal} {Phys. Rev. B}\ }\textbf {\bibinfo {volume} {113}},\ \bibinfo
  {pages} {L060403} (\bibinfo {year} {2026}{\natexlab{d}})}\BibitemShut
  {NoStop}%
\bibitem [{\citenamefont {Yarmohammadi}\ and\ \citenamefont
  {Cheraghchi}(2020)}]{PhysRevB.102.075411}%
  \BibitemOpen
  \bibfield  {author} {\bibinfo {author} {\bibfnamefont {M.}~\bibnamefont
  {Yarmohammadi}}\ and\ \bibinfo {author} {\bibfnamefont {H.}~\bibnamefont
  {Cheraghchi}},\ }\bibfield  {title} {\bibinfo {title} {Effective low-energy
  {RKKY} interaction in doped topological crystalline insulators},\ }\href
  {https://doi.org/10.1103/PhysRevB.102.075411} {\bibfield  {journal} {\bibinfo
   {journal} {Phys. Rev. B}\ }\textbf {\bibinfo {volume} {102}},\ \bibinfo
  {pages} {075411} (\bibinfo {year} {2020})}\BibitemShut {NoStop}%
\bibitem [{\citenamefont {Cheraghchi}\ and\ \citenamefont
  {Yarmohammadi}(2021)}]{Cheraghchi2021}%
  \BibitemOpen
  \bibfield  {author} {\bibinfo {author} {\bibfnamefont {H.}~\bibnamefont
  {Cheraghchi}}\ and\ \bibinfo {author} {\bibfnamefont {M.}~\bibnamefont
  {Yarmohammadi}},\ }\bibfield  {title} {\bibinfo {title} {Anisotropic
  ferroelectric distortion effects on the {RKKY} interaction in topological
  crystalline insulators},\ }\href {https://doi.org/10.1038/s41598-021-84398-0}
  {\bibfield  {journal} {\bibinfo  {journal} {Scientific Reports}\ }\textbf
  {\bibinfo {volume} {11}},\ \bibinfo {pages} {5273} (\bibinfo {year}
  {2021})}\BibitemShut {NoStop}%
\bibitem [{\citenamefont {Ke}\ \emph {et~al.}(2024)\citenamefont {Ke},
  \citenamefont {Asmar},\ and\ \citenamefont {Tse}}]{PhysRevB.110.035307}%
  \BibitemOpen
  \bibfield  {author} {\bibinfo {author} {\bibfnamefont {M.}~\bibnamefont
  {Ke}}, \bibinfo {author} {\bibfnamefont {M.~M.}\ \bibnamefont {Asmar}},\ and\
  \bibinfo {author} {\bibfnamefont {W.-K.}\ \bibnamefont {Tse}},\ }\bibfield
  {title} {\bibinfo {title} {{F}loquet-driven indirect exchange interaction
  mediated by topological insulator surface states},\ }\href
  {https://doi.org/10.1103/PhysRevB.110.035307} {\bibfield  {journal} {\bibinfo
   {journal} {Phys. Rev. B}\ }\textbf {\bibinfo {volume} {110}},\ \bibinfo
  {pages} {035307} (\bibinfo {year} {2024})}\BibitemShut {NoStop}%
\bibitem [{\citenamefont {Lindemann}(1910)}]{Lindemann1910}%
  \BibitemOpen
  \bibfield  {author} {\bibinfo {author} {\bibfnamefont {F.~A.}\ \bibnamefont
  {Lindemann}},\ }\bibfield  {title} {\bibinfo {title} {The calculation of
  molecular vibration frequencies},\ }\href
  {https://ntrs.nasa.gov/citations/19840027015} {\bibfield  {journal} {\bibinfo
   {journal} {Phys. Z.}\ }\textbf {\bibinfo {volume} {11}},\ \bibinfo {pages}
  {609} (\bibinfo {year} {1910})}\BibitemShut {NoStop}%
\end{thebibliography}%
		
	\end{document}